\documentstyle[12pt,axodraw]{article}

\newcommand{\gsim}{\raisebox{-0.13cm}{~\shortstack{$>$ \\[-0.07cm] $\sim$}}~}

\newcommand{\ra}{\rightarrow}
\newcommand{\ee}{e^+e^-}
\newcommand{\tb}{\tan \beta}
\newcommand{\s}{\smallskip}
\newcommand{\nn}{\noindent}
\newcommand{\non}{\nonumber}
\newcommand{\beq}{\begin{eqnarray}}
\newcommand{\eeq}{\end{eqnarray}}

\catcode`@=11
\def\citer{\@ifnextchar
[{\@tempswatrue\@citexr}{\@tempswafalse\@citexr[]}}

\def\@citexr[#1]#2{\if@filesw\immediate\write\@auxout{\string\citation{#2}}\fi
  \def\@citea{}\@cite{\@for\@citeb:=#2\do
    {\@citea\def\@citea{--\penalty\@m}\@ifundefined
       {b@\@citeb}{{\bf ?}\@warning
       {Citation `\@citeb' on page \thepage \space undefined}}%
\hbox{\csname b@\@citeb\endcsname}}}{#1}}
\catcode`@=12

\begin{document}

\baselineskip=16pt
\renewcommand{\thefootnote}{*}

\begin{flushright}
CERN TH/2003--252\\
PM/03--22\\
PSI--PR--03--17\\
November 2003\\
\end{flushright}

\vspace*{0.5cm}

\begin{center}

{\Large\sc {\bf {\tt SDECAY}: a Fortran code for the decays of}}

\vspace*{0.2cm}

{\Large\sc {\bf the supersymmetric particles in the MSSM\footnote{The code can
be obtained at the url: http://people.web.psi.ch/muehlleitner/SDECAY}}}

\vspace{0.7cm}

{\large\sc M. M\"uhlleitner$^1$}, {\large\sc A. Djouadi$^{2,3}$} and 
{\large\sc Y. Mambrini$^4$}

\vspace{0.7cm}

$^1$ Paul  Scherrer Institute, CH--5232 Villigen PSI, Switzerland. 
\vspace*{2mm}

$^2$ Theory Division, CERN, CH--1211 Geneva 23, Switzerland. 
\vspace*{2mm} 

$^3$ Laboratoire de Physique Math\'ematique et Th\'eorique, UMR5825--CNRS,\\
Universit\'e de Montpellier II, F--34095 Montpellier Cedex 5, France. 
\vspace*{2mm}

$^4$ Laboratoire de Physique Th\'eorique, UMR8627--CNRS, \\
Universit\'e Paris-Sud, F--91405 Orsay, France. 

\end{center} 

\vspace*{.5cm} 

\begin{abstract}  
\nn We present the Fortran code {\tt SDECAY}, which calculates the decay
widths  and branching ratios of all the supersymmetric particles in the
Minimal  Supersymmetric Standard Model, including higher order effects. Besides
the usual two-body decays of fermions and gauginos and the three-body decays of
charginos, neutralinos and gluinos, we have also implemented the three-body
decays of top squarks, and  even the four-body decays of the top squark; 
the important loop-induced decay modes are also included. The  QCD
corrections to the two-body decays involving strongly interacting particles
and the dominant components of the electroweak corrections to all decay modes
are implemented.   

\end{abstract}

\newpage 
\setcounter{footnote}{0}
\renewcommand{\thefootnote}{\arabic{footnote}}

\subsection*{1. Introduction}

The search for the new particles predicted by supersymmetric theories is a
major goal of present and future colliders. A large theoretical effort has been
devoted in the last two decades to determine the basic properties of these
particles, as well as their decay modes and production mechanisms in collider
experiments. Most of these studies have been  performed in the framework of the
Minimal Supersymmetric Standard Model (MSSM) \citer{MSSM1,MSSM3},
which has the minimal gauge structure, particle content and minimal
interactions, and leads to a stable lightest supersymmetric particle (LSP). \s 

As is well known, it is a very complicated task to deal in an exhaustive way
with all the properties of these new particles  and to make detailed and
complete  phenomenological analyses  and comparisons with the outcome of or
expectations  from experiments. This is mainly due to the fact that, even in 
the MSSM, there are more than a hundred new parameters in the most general
case, and even if one constrains the model to have a viable  phenomenology,
there are still over 20 free parameters left to cope with.  This large number
of inputs enters in the evaluation of the masses of ${\cal O} (30)$ 
supersymmetric (SUSY)
particles and Higgs bosons as well as their complicated couplings,  which
involve several non-trivial aspects, such as the mixing between different
states, the Majorana nature of some particles and, if the aim is to be quite
precise, the higher order corrections.  One then has to calculate, in the most
accurate way, the rates for the many possible decay modes and production
processes at the  various possible machines. \s

Once SUSY particles are found, their properties are expected to be
determined with an accuracy of a few per cent at the LHC \cite{Houches} and a
precision at the per cent level or below at future $\ee$ linear colliders
\cite{TESLA}.  To match this expected experimental accuracy, we need to
calculate the mass  spectra, the various couplings, the decay  branching ratios
and the production cross sections with a rather high precision, i.e. including
the higher order effects. Leaving aside the production processes which are
dealt with by Monte Carlo event generators, and for which the cross sections
have been calculated at next-to-leading order (NLO) in some cases
\cite{NLOprod}, one therefore needs to achieve the following goals: \s

-- The physical (pole) masses and the various couplings [as well as the soft
SUSY-breaking parameters which enter those] of the SUSY particles and the MSSM
Higgs bosons  need to be calculated  very accurately. They  must include all
relevant features such as the masses of third-generation fermions, the mixing
between the various states, and the radiative corrections when important. In
constrained MSSMs two additional features need to be handled carefully: the 
renormalization group evolution (RGE) of parameters between the low-energy
scale and  the high-energy scale and the consistent implementation of radiative
electroweak symmetry breaking (EWSB), i.e. the loop corrections to the
effective scalar potential.  There are  several available RGE codes 
\citer{isajet,spheno}, which calculate the supersymmetric particle and Higgs
boson masses as well as the soft SUSY-breaking parameters in the unconstrained
and constrained MSSMs; they can be straightforwardly extended to allow for the
calculation of the various and numerous couplings. \s 

-- All the possible two-body decay modes that can occur at the tree level
\cite{2body} should be taken into account. These consist  not only of decays of
the inos\footnote{For simplicity, we will collectively call inos the charginos
and neutralinos, and sometimes the gluinos.} into fermion/sfermion pairs or
sfermion decays into fermion/ino pairs, but also decays involving Higgs and
gauge bosons in the final state. The QCD corrections, which are known to be
rather large \citer{QCD1,QCD3}, need to be incorporated in the processes
involving strongly interacting particles. In addition, some electroweak
radiative corrections can be as large as the QCD corrections, and in principle,
they should be taken into account. However, it is a tremendous task to
calculate all these corrections, and they are available in the one-loop
approximation \cite{NLOEWd} only for a very limited number of processes.
Nevertheless, the bulk of these corrections, those stemming from the running of
the gauge and Yukawa  couplings and the running of some soft SUSY-breaking
parameters, is available in the literature and can be incorporated with a
minimum of effort. \s

-- All possibly important higher order decay modes must be included. They
consist of the three-body decay modes of the charginos, neutralinos and the 
gluino into lighter inos  and two massless fermions  \cite{3body1,3body2},
which are known to be important, 
but also of the three-body decays of third-generation sfermions,
which have been shown recently to be possibly important in some kinematical and
parameter configurations \cite{3body3,3body4}. This is particularly the case
for the top squark, when its mass is smaller than the lighter chargino mass and
the sum of the masses of the lightest  neutralino and the top quark. In fact,
even the four-body decay mode of the top squark \cite{4body} has been shown to
possibly compete with the flavour changing neutral current (FCNC) and 
loop-induced decay into a charm quark and
the lightest neutralino \cite{loop1}, which also has to be included. Another
set of loop-induced  decay modes  which might play a prominent role are the
radiative decays of the  next-to-lightest neutralino into the lightest
neutralino and a photon \cite{loop2,loop3} and, to a lesser extent, the decay
of the gluino into a gluon and the LSP \cite{gluinorad}.\s

The Fortran code {\tt SDECAY}, which we present in this report, deals with the
decays of SUSY particles in the framework of the MSSM, and includes the most
important higher order effects. It uses the RGE program {\tt SuSpect}
\cite{suspect} for the calculation of the mass spectrum and the soft
SUSY-breaking parameters [but of course, the program can be easily linked to
any other RGE code] and evaluates the various couplings of the SUSY particles
and MSSM Higgs bosons.  It calculates the decay widths and the branching ratios
of all the two-body decay modes, including the QCD corrections to the processes
involving squarks and gluinos and the dominant electroweak effects to all
processes. It also calculates the loop-induced  two-body decay channels, as
well as all the  possibly important higher order decay modes: the three-body
decays of charginos, neutralinos, gluinos and third generation sfermions and
the four-body decays of the top squark\footnote{There are other programs which
calculate the decay branching ratios of SUSY particles \cite{isajet,spheno,
susygen},
and that also include some higher order effects. For instance, the loop decays
and the three-body ino decays are included in {\tt ISASUGRA}, {\tt SPHENO} and
{\tt SUSYGEN}; the program {\tt SPHENO} also deals with some three-body decays
of the top squark. None of these programs includes the QCD corrections and the
full set of higher order decays of third-generation sfermions, though.}. \s

The program also calculates the decay widths and branching ratios of the 
heavy top quark. Besides the standard decay into a $W$ boson and a bottom
quark, the top quark decay widths into a charged Higgs boson and a bottom
quark and into a top squark and a neutralino, are evaluated. 
The one-loop SUSY-QCD radiative corrections that are known \cite{QCDtop} 
will be included in an upgraded version of the program. \s

Besides the {\tt SuSpect} files needed for the evaluation of the spectrum, the
code contains only one source file {\tt sdecay.f}, written in Fortran77, and
one input file, {\tt sdecay.in}, from which any choice of approximation in the
calculation is driven [including or not the higher order corrections and/or 
decays, the choice of the various scales, the order of perturbation at which
some couplings are calculated, etc.]. All results for the total decay widths
and branching ratios are given in the output file {\tt sdecay.out}, either in a
simple form  or in the SUSY Les Houches Accord (SLHA)  \cite{SLHA} form [this
choice can be made in the input file].  The program is very user-friendly, 
self-contained and it can easily be linked with other codes or Monte Carlo
event generators. It is rather fast and flexible, thus allowing scans of   the
parameter space with several possible options and choices  for model
assumptions and approximations. \s

The program {\tt SDECAY} is of the same level of sophistication as the program
{\tt HDECAY} \cite{hdecay},  which calculates the decay widths and the
branching ratios of the [Standard Model and] MSSM Higgs bosons, including  all
relevant higher order effects. In fact, the three programs {\tt SuSpect}, {\tt
HDECAY} and {\tt SDECAY}, have many common features and subroutines, and are
organized in a similar way. They provide a coherent, consistent and 
comprehensive description of the properties of the supersymmetric and Higgs
particles in the MSSM, prior to the level of production  which, as was
mentioned previously, is the domain or ``chasse gard\'ee" of the Monte Carlo 
event  generators. A light version, which combines these three programs, and
which can be easily linked to any Monte Carlo  generator, is under development
and will appear quite soon \cite{SSH}. \s

The rest of this report is organized as follows. In the next section, we will
summarize the main features of the MSSM that we will deal with, concerning the 
sparticle and Higgs boson masses and couplings, and the notation that we use in
the program.  In section 3, we discuss all the decay modes that are implemented
and the way the higher order decays and the radiative corrections are included.
In section 4,  we discuss the main features and the structure of the program, 
briefly summarize how it works, and display the content of the input and output
files. In section 5,  a short conclusion will be given.

\subsection*{2. The implementation of the MSSM}

The program {\tt SDECAY} deals with the MSSM, i.e. with the basic assumptions 
of a:

\begin{itemize}
\vspace*{-2mm}

\item[--] minimal gauge group, the Standard Model ${\rm SU(3)_C \times SU(2)_L 
\times U(1)_Y}$ one, 
\vspace*{-2mm}

\item[--] minimal particle content: three generations of ``chiral" sfermions 
$\tilde{f}^i_{L,R}$ [no right-handed sneutrinos] and two 
doublets of Higgs fields $H_1$ and $H_2$,
 \vspace*{-2mm}

\item[--] minimal set of couplings imposed by R-parity conservation to enforce 
baryon and lepton number conservation in a simple way  and which leads to
a stable LSP, 
\vspace*{-2mm}

\item[--] minimal set of soft SUSY-breaking parameters: gaugino mass terms 
$M_i$, scalar mass terms $m_{H_i}$ and $m_{\tilde{f}_i}$, a 
bilinear term $B$ and trilinear sfermion couplings $A_i$.  
\vspace*{-2mm}
\end{itemize}

For the superpotential and the minimal set of soft SUSY-breaking, which give
all the interactions and couplings, we follow the notations that can be found
in the users manual of the program {\tt SuSpect} \cite{suspect}, which we  use
for the determination of the SUSY particle and Higgs boson spectra. To have a
viable phenomenology and a reduced number of free parameters, we thus also make
the following three assumptions: 

\begin{itemize}
\vspace*{-2mm}

\item[$(i)$] All the soft SUSY-breaking parameters are real and therefore 
no new source of CP-violation is generated, in addition to the one from
the CKM  matrix.
\vspace*{-2mm}

\item[$(ii)$] The matrices for the sfermion masses and for the trilinear
couplings  are all diagonal,  implying the absence of FCNCs at the
tree level.  
\vspace*{-2mm}

\item[$(iii)$] The first and second sfermion generations are universal at low 
energy to  cope with some severe experimental constraints [this is also  
motivated by the fact that we have neglected for simplicity all the masses of 
 the first- and second-generation fermions which are small enough not to have 
any  significant effect]. 
\vspace*{-2mm}
\end{itemize}

Making these three assumptions will lead to the so-called ``phenomenological
MSSM" (or pMSSM) discussed in \cite{MSSM3}, with 22 input parameters only: \s

\nn \hspace*{2cm} $\tan \beta$: the ratio of the {\it vev}s of the two-Higgs 
doublet fields.\\
 \hspace*{2cm} $m^2_{H_1}, m^2_{H_2}$: the Higgs mass parameters squared. \\
 \hspace*{2cm} $M_1, M_2, M_3$: the bino, wino and gluino mass parameters. \\
 \hspace*{2cm} $m_{\tilde{q}}, m_{\tilde{u}_R}, m_{\tilde{d}_R}, 
               m_{\tilde{l}}, m_{\tilde{e}_R}$: the first/second-generation
 sfermion mass parameters.\\ 
  \hspace*{2cm} $m_{\tilde{Q}}, m_{\tilde{t}_R}, m_{\tilde{b}_R}, 
               m_{\tilde{L}}, m_{\tilde{\tau}_R}$: the third-generation
 sfermion mass parameters.\\
  \hspace*{2cm} $A_u, A_d, A_e$: the first/second-generation trilinear 
  couplings. \\
  \hspace*{2cm} $A_t, A_b, A_\tau$: the third-generation trilinear couplings. \s

If one requires a proper electroweak symmetry breaking, the 
Higgs-higgsino (supersymmetric) mass parameter $|\mu|$ (up to a sign) and the
soft SUSY-breaking  bilinear Higgs term $B$ are determined, given the above
parameters [alternatively, one can trade  the values of $m^2_{H_1}$ and
$m^2_{H_2}$ with the ``more physical" pseudoscalar Higgs boson  mass $M_A$ and
parameter $\mu$, a possibility provided by the program {\tt SuSpect}, which
deals with all the aspects of EWSB]. \s

In constrained models, such as minimal Supergravity (mSUGRA) \cite{mSUGRA},
the gauge-mediated SUSY breaking (GMSB) \cite{GMSB} and anomaly-mediated SUSY
breaking  (AMSB) \cite{AMSB} models, most of the 22 soft SUSY-breaking input
parameters of the pMSSM listed above are derived from a set of universal
boundary conditions at the high-energy scale [the Grand Unification (GUT) scale
for the mSUGRA and AMSB models and the messenger scale for the GMSB model]. In
the mSUGRA case, for instance, the entire set of soft SUSY-breaking parameters
[and thus the superparticle and Higgs spectrum] is determined by the values of 
only five free parameters: a common gaugino mass $m_{1/2} = M_i$, a universal 
scalar mass $m_0= m_{\tilde{f}}= m_{H_i}$ and a universal trilinear coupling
$A_0=A_i$  at the GUT scale, the sign of the higgsino parameter $\mu$ and
$\tb$. \s

The low-energy soft SUSY-breaking parameters are then derived  from the
high-energy  ones above through Renormalization Group Equations. [Note
that the values of $|\mu|$ and $M_A$ are obtained by requiring proper EWSB, 
which
should be implemented, and the value of $M_{\rm GUT}$ is defined as the scale
where the three gauge coupling constants of the Standard Model (SM) unify]. 
One then proceeds to
calculate the pole masses of the Higgs bosons and all the supersymmetric
particles, including the possible mixing between the current  states and the
radiative corrections [up to two loops in some cases] when they are
important.\s

All these steps are performed by the program {\tt SuSpect} and, for completeness,
we reproduce in Fig.~1  the general algorithm that is used in the code
\cite{suspect}. This iterative  algorithm  includes the various important 
steps of the calculation:  the choice of SM input parameters at low energy
[the gauge coupling constants and the pole masses of the third-generation
fermions], the calculation of the running couplings including radiative
corrections in the modified  Dimensional Reduction $\overline{\rm DR}$ scheme
[which preserves SUSY] and  their RG running back and forth between the low and
high scales, with the  possibility of imposing the unification of the gauge
couplings and the inclusion of SUSY thresholds in some cases, the RG evolution
of the soft SUSY-breaking parameters from the high scale to the EWSB scale,
the minimization of  the one-loop effective potential and the determination of
some important parameters, and finally the calculation of the particle masses
including the  diagonalization of the mass matrices and the radiative
corrections.\s 

To be more specific, we provide below a summary list of the higher order
effects that have been included in the program: \s

-- For the $\overline{\rm DR}$ gauge and third-generation fermion Yukawa
couplings, defined at the scale $M_Z$, the full set of standard and SUSY
corrections has been implemented according to the approach of  Pierce, Bagger,
Matchev and Zhang (PBMZ) \cite{PBMZ}. There are two exceptions: in the case of
$\sin^2\theta_W$, the small SUSY particle  contributions to the box diagrams
have been omitted, and in the case of $m_b$ and $m_\tau$, only the QCD and the
leading electroweak corrections at zero-momentum transfer have  been
incorporated [which, according to PBMZ, is a very good approximation].  In some
cases,  again according to the PBMZ approach,  some important two-loop
corrections have also been taken into account. \s

\vspace*{3mm}
\begin{picture}(1000,470)(10,0)
\Line(390,450)(450,450)
\ArrowLine(450,-60)(450,450)
\Line(450,-60)(390,-60)

\Line(10,470)(390,470)
\Line(10,410)(390,410)
\Line(10,470)(10,410)
\Line(390,470)(390,410)

\Text(200,460)[]{Choice of low energy input:  $\alpha(M_Z), \sin^2\theta_W, 
\alpha_S(M_Z)$, $m_{t,b,\tau}^{\rm pole}$\, ; $\tan \beta (M_Z)$}

\Text(200,440)[]{Radiative corrections $\Rightarrow$ $g_{1,2,3}^{\rm 
\overline{DR}}(M_Z)$, $\lambda_\tau^{\rm \overline{DR}} (M_Z), \lambda_b^{\rm 
\overline{DR}}(M_Z), \lambda_t^{\rm \overline{DR}} (M_Z)$}

\Text(200,420)[]{\it First iteration: no SUSY radiative corrections.} 

\ArrowLine(200,408)(200,392)

\Line(10,390)(390,390)
\Line(10,310)(390,310)
\Line(10,390)(10,310)
\Line(390,390)(390,310)

\Text(200,370)[]{Two--loop RGE for $g_{1,2,3}^{\overline{\rm DR}}$ and 
$\lambda_{\tau,b,t}^{\overline{\rm DR}}$ with choice: $\begin{array}{l}  
g_1=g_2 \cdot \sqrt{3/5} \\  M_{\rm GUT} \sim 2 \times 10^{16}~{\rm  GeV} 
\end{array}$}

\Text(200,340)[]{Include all SUSY thresholds via step functions in $\beta$ 
functions.}

\Text(200,320)[]{\it First iteration: unique threshold guessed.} 

\ArrowLine(200,308)(200,295)

\Text(200,285)[]{Choice of SUSY-breaking model (mSUGRA, GMSB, AMSB, or pMSSM).}
 
\Text(200,265)[]{Fix your high--energy input (mSUGRA: $m_0, m_{1/2}, A_0$, 
sign($\mu)$, etc.).} 

\ArrowLine(200,253)(200,242)

\Line(10,240)(390,240)
\Line(10,180)(390,180)
\Line(10,240)(10,180)
\Line(390,240)(390,180)

\Text(200,220)[]{Run down with RGE to: $\begin{array}{l} - M_Z~{\rm for}~
g_{1,2,3}~{\rm and}~\lambda_{\tau,b,t} \\
- M_{\rm EWSB}~{\rm for}~\tilde{m}_i, M_i, A_i, \mu, B \end{array}$} 
\Text(200,195)[]{\it First iteration: guess for $M_{\rm EWSB} = M_Z$.}

\ArrowLine(200,178)(200,162)

\Line(10,160)(390,160)
\Line(10,100)(390,100)
\Line(10,160)(10,100)
\Line(390,160)(390,100)

\Text(200,150)[]{$\mu^2, \mu B = F_{\rm non-linear}(m_{H_1}, m_{H_2}, 
\tan\beta, V_{\rm loop} )$}

\Text(200,130)[]{$V_{\rm loop} \equiv $ Effective potential at one loop with 
all masses.}

\Text(200,110)[]{{\it First iteration: $V_{loop}$ not included} }

\Line(390,130)(420,130)
\ArrowLine(420,30)(420,130)
\Line(420,30)(390,30)

\ArrowLine(200,98)(200,90)

\Text(200,85)[]{Check of consistent EWSB ($\mu$ convergence, no tachyons, 
simple CCB/UFB, etc.) } 

\ArrowLine(200,75)(200,62)

\Line(10,60)(390,60)
\Line(10,0)(390,0)
\Line(10,60)(10,0)
\Line(390,60)(390,0)

\Text(200,50)[]{Diagonalization of mass matrices and calculation of masses /  
couplings}

\Text(200,30)[]{Radiative corrections to the physical Higgs, sfermion, gaugino
masses.}

\Text(200,10)[]{\it First iteration: no radiative corrections.} 

\ArrowLine(200,-2)(200,-17)

\Line(10,-20)(390,-20)
\Line(10,-85)(390,-85)
\Line(10,-20)(10,-85)
\Line(390,-20)(390,-85)

\Text(200,-30)[]{Check of a reasonable spectrum:} 

\Text(200,-45)[]{-- no tachyonic masses (from RGE, EWSB or mix), 
good LSP, etc..\ \hfill } 

\Text(200,-60)[]{-- not too much fine-tuning and sophisticated CCB/UFB 
conditions,\hfill }

\Text(200,-75)[]{-- agreement with experiment: $\Delta\rho$, $(g-2)$, 
$b \to s\gamma$. \ \ \ \ \ \ \ \ \ \ \ \ \ \ \ \ \ \ \ \ \hfill}
\end{picture}

\vspace*{3.3cm}

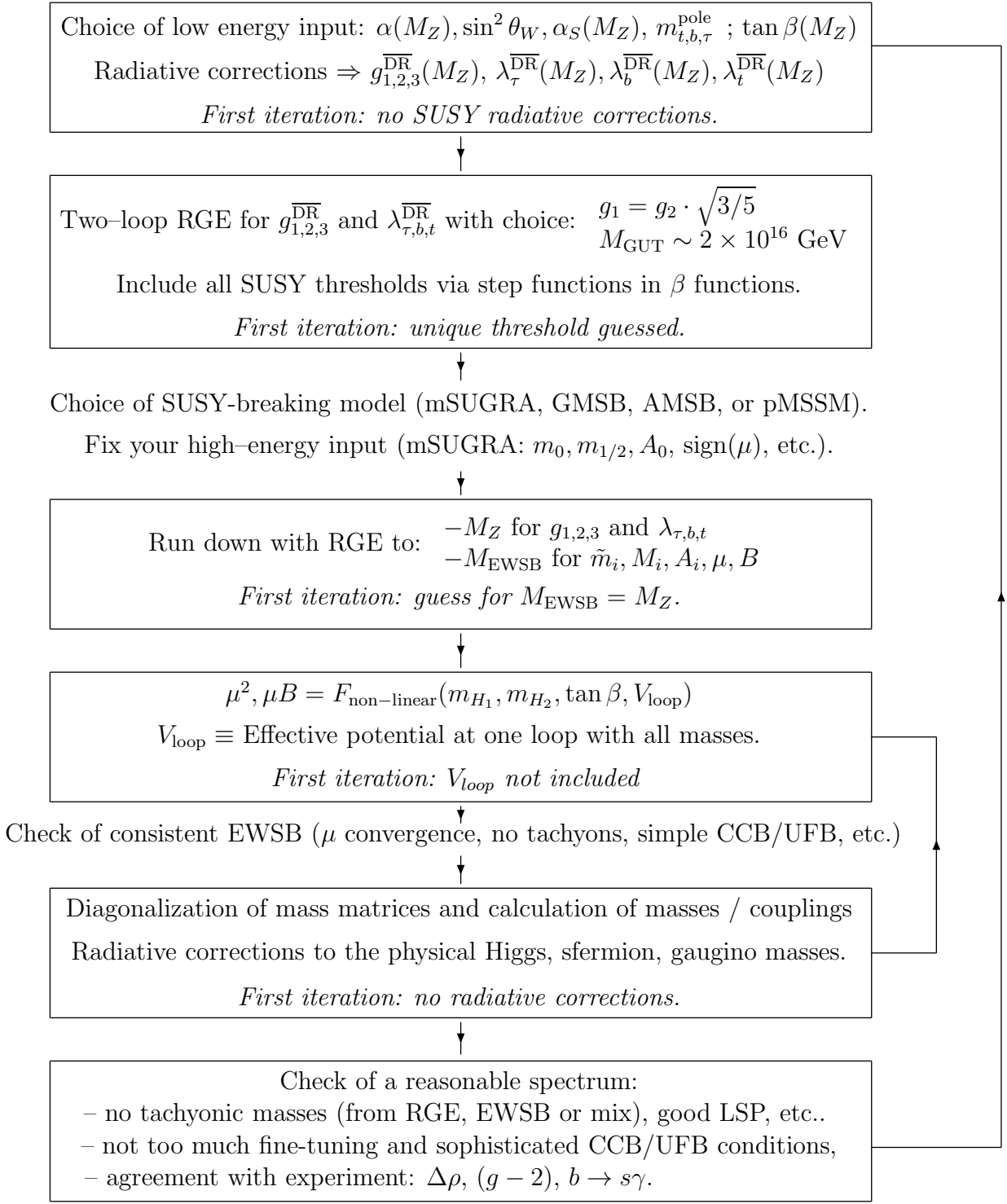
\noindent Figure 1: {\it Iterative algorithm for the calculation of the SUSY
particle spectrum in {\tt SuSpect} from the choice of input parameters (first
step) to the check of the  spectrum (last step). The EWSB ``small" iteration 
on $\mu$  the RG/RC ``long" iteration are performed until a satisfactory
convergence is reached.}

\setcounter{figure}{1}

-- The RGEs have been used at the two-loop level for the gauge and Yukawa
couplings, as well as for the three gaugino mass parameters $M_1,M_2,M_3$ and 
the two Higgs mass parameters $m_{H_1},m_{H_2}$. For the other soft
SUSY-breaking parameters [essentially the sfermion mass parameters
$m_{\tilde{f}}$ and the trilinear couplings $A_f$], only the one-loop RGEs have
been used. The GUT scale $M_{\rm GUT}$ can be either defined as the value at
which the two gauge couplings $g_1$ and $g_2$ unify, or can be set by hand
at $M_{\rm GUT} \sim 2 \times 10^{16}~{\rm  GeV}$. \s

-- The EWSB has been implemented through the tadpole method. The full one-loop
standard and SUSY contributions to the tadpoles have been taken into account.
The dominant  two-loop corrections, those stemming from QCD and the 
third-generation fermion Yukawa couplings, have also been  included. The EWSB
scale has been chosen to be the geometric mean of the two top squark
masses,  $M_{\rm EWSB} = (m_{\tilde{t}_1} m_{\tilde{t}_2})^ {1/2}$. \s

-- The soft SUSY-breaking parameters and the parameter $\mu$ that we obtain
are all $\overline{\rm DR}$ parameters defined at the scale $M_{\rm EWSB}$. 
Using these parameters, the chargino and neutralino mass matrices are 
diagonalized with real matrices $U/V$ and $Z$, respectively, to obtain the 
tree-level physical ino masses.  The third-generation sfermion mass matrices
are also diagonalized to obtain the tree-level sfermion masses,  the mixing
angles $\theta_{\tilde{t}}, \theta_{\tilde{b}}$ and $\theta_{\tilde{\tau}}$
being $\overline{\rm DR}$ parameters defined at the scale $M_{\rm EWSB}$. \s

-- The radiative corrections to the sfermion masses are again included
according to PBMZ \cite{PBMZ}, i.e. only the QCD corrections for the
superpartners of light quarks [including the bottom squark] plus the full QCD 
and electroweak corrections to the two top squarks; the small electroweak
radiative corrections to the slepton masses [which according to PBMZ are at the
level of one per cent] have been neglected. The full one-loop QCD radiative
corrections to the gluino mass are incorporated, while in the
chargino/neutralino case, the radiative corrections to the masses are simply
included in the gaugino and higgsino limits, which is a very good approximation
according to PBMZ. \s

-- The calculation of the masses of the MSSM Higgs bosons and the mixing angle
$\alpha$ in the CP-even sector can be made using four routines that are
available on the market: {\tt Subhpole}  \cite{SUBH}, {\tt HMSUSY} \cite{HHH}, 
{\tt FeynHiggsFast} \cite{FeynHiggsFast} and {\tt Hmasses} \cite{BDSZ}. In view
of the approximations used in {\tt SuSpect}, it is more appropriate to use the
routine {\tt Hmasses}, which has the same level of approximation, since it
includes the full one-loop radiative corrections, and the dominant two-loop
corrections at order $\alpha_s \lambda_f^2$ and $\lambda_f^4$, $\lambda_f$ 
being the Yukawa couplings of the third-generation fermions, $f=t,b$ and
$\tau$.\s

Using the gauge couplings, the third-generation fermion masses and the soft
SUSY-breaking parameters discussed above, we then proceed in the program {\tt
SDECAY} to calculate all the couplings  of the SUSY particles and the MSSM
Higgs bosons. In most cases, we use the Feynman rules given by Haber et al. in
Ref.~\cite{MSSM2}.  These couplings are thus defined in the $\overline{\rm DR}$
scheme and evaluated at the scale $M_{\rm EWSB}$. Therefore, they already
include some radiative corrections, and care should be taken when they are used
in one-loop-corrected amplitudes  to avoid double counting, as will be
discussed later. 

\subsection*{3. Decays of the SUSY particles}

\subsubsection*{3.1 Two-body decays at the tree level}

\hspace*{.5cm} \underline{The main decay modes of sfermions}  will be into
their partner fermions and neutralinos, as well as into their isospin partner
fermions and charginos\footnote{Here and in the following, we  collectively
denote by  $\chi_i$ the charginos and neutralinos, and we discard the
distinction between the two isospin sfermion partners.}
\beq \tilde{f}_i & \ra & f \chi_j \ .  \eeq
In the case of squarks, when they are heavier than the gluino, they can also 
decay into gluino-quark final states
\beq
\tilde{q}_i \ra q \tilde{g} \ .
\eeq
If the mass splitting between two sfermions of the same generation is large
enough, as can be the case of the third-generation $(\tilde{t},\tilde{b})$ 
and $({\tilde\nu_\tau}, \tilde{\tau})$ isodoublets, the heavier sfermion can
decay into the lighter one plus a gauge boson $V=W,Z$  or a Higgs boson
$\Phi=h,H,A,H^\pm$
\beq
\tilde{f}_i & \to & \tilde{f}_j' V  \ , \\
\tilde{f}_i & \to & \tilde{f}_j' \Phi \ .
\eeq

\underline{The heavier neutralinos and charginos}  will decay into the lighter 
chargino and neutralino states and gauge or Higgs bosons 
\beq
\chi_i &\ra& \chi_j V \\
\chi_i &\ra& \chi_j \Phi
\eeq
and, if enough phase space is available, into fermion-sfermion pairs
\beq
\chi_i \ra f  \tilde{f}_j \ .
\eeq
For gluinos, when they are heavier than squarks, their only relevant 
decay channel will be into quark plus squark final states: 
\beq
\tilde{g} \to q \tilde{q} \ .
\eeq

\underline{In the case of the GMSB model}, the lightest SUSY particle is
the gravitino $\tilde{G}$. The next-to-lightest SUSY particle (NLSP) can be
either the  lightest neutralino or the lightest sfermion, in general the
$\tilde{\tau}_1$ [in which case, the decay $\chi_1^0 \to \tilde{\ell} \ell$ is
allowed kinematically]. If the NLSP is a slepton, its only allowed decay is
into a lepton and a gravitino, $\tilde{\ell} \to \ell \tilde{G}$, with a
branching ratio of 1. If the NLSP is the lightest neutralino, there are
several possible decays, $\chi_1^0 \to \gamma \tilde{G}, Z \tilde{G}$ and 
$\tilde{G} \Phi$ with $\Phi=h,H,A$. \s

The program {\tt SDECAY} calculates the partial widths and the branching ratios
of all these decays, including all possible combinations. It first checks if
the decay is phase-space-allowed and then uses the two-body simple formulae
available in the literature; see for instance Refs.~\cite{3body2,3body4}. The
masses involved in the phase space are all pole masses [i.e., in the case of
the SUSY and Higgs particles, the one-loop renormalized masses given by {\tt
SuSpect}].  The masses of the third-generation fermions [the only ones in the
program that are assumed to be non-zero] are the on-shell masses when they
enter the phase space but the running $\overline{\rm DR}$ masses defined at the
scale $M_{\rm EWSB}$ when they enter the various couplings. This is also the
case of all the soft SUSY-breaking  parameters and the third-generation
sfermion mixing angles that enter the couplings. Nevertheless, we have left as
an option the possibility for the QCD coupling constant and the $b,t$  Yukawa
couplings, to be evaluated at the scale of the decaying superparticle or any
other scale. However, in this case, only the standard QCD corrections are
included in the running, following the approach of Ref.~\cite{QCDrunning}.   

\subsubsection*{3.2 QCD corrections to the two-body decays}

The one-loop QCD corrections have been incorporated to the two-body decay 
processes involving (s)quarks or gluinos in the initial or final state; more 
specifically, they have been included in the following processes
\beq
\tilde{q}_i &\to& q \chi_j   \\
\tilde{q}_i &\to& \tilde{q}_j \Phi \\
\tilde{q}_i &\to& q \tilde{g} \ \ \ {\rm and} \ \tilde{g} \to \tilde{q}_i q \ .
\eeq

We have used the formulae given in Refs.~\cite{QCD1}, \cite{QCD2} and 
\cite{QCD3} for, respectively, the processes of eqs.~(9), (10) and (11). 
The corrections for the process $\chi_i \to \tilde{q}_i q$ that can be 
adapted from the ones of the reverse process, eq.~(9), and the corrections
to the decay $\tilde{q}_i \to \tilde{q}_j V$ will be implemented in the 
next version of the program. The Passarino-Veltman one-, two- and three-point
functions for the loop amplitudes have been implemented using the formulae
given in  Ref.~\cite{PV}.  For the three-body phase-space integrals, when an
additional gluon is emitted in the final state, we use the analytical 
formulae given in Refs.~\citer{QCD1,QCD3} and which involve at most Spence
functions. \s

A few remarks are worth making at this stage: \s

$(i)$ All the corrections have been incorporated in the $\overline{\rm DR}$
scheme, while in the previous references some corrections were calculated
in the $\overline{\rm MS}$ scheme. In the latter case the differences 
between the two schemes have to be corrected by additional counterterms.
The results for the physical observables of course will be the same in the two 
schemes up to the calculated order, as it should be. \s

$(ii)$ All the masses of the  particles involved in the processes, in
particular those of the strongly interacting particles, which  in principle
need to be renormalized, are pole masses, i.e. are defined on the mass shell.
The self-energies of these particles are defined in such a way that the
residues at the poles are unity.  \s

$(iii)$ Because the top and bottom quark masses and the stop and sbottom mixing
angles [and thus the trilinear couplings $A_t$ and $A_b$] that are obtained
using the program {\tt SuSpect}, already include some one-loop contributions, 
some care has to be taken when dealing with the renormalization of these
parameters and their one-loop counterterms to avoid double counting [in
fact only the divergent pieces have to be included]. \s

$(iv)$ In the case of decays involving gluinos, which are strong interaction
decays already at the tree level, a special treatment is needed for $\alpha_s
(\mu_R)$, where $\mu_R$ is the renormalization scale. The heavy particles, 
top quarks, squarks and gluinos should be removed from the $\mu^2_R$ evolution
of $\alpha_s$  and decoupled for momenta smaller than their masses.  \s

Finally, as mentioned previously, the bulk of the electroweak radiative
corrections which is due to the running of the gauge and third-generation
fermion Yukawa couplings, is already taken into account since these parameters,
when appearing in the amplitudes, have been evaluated at the EWSB scale. The
remaining corrections [including photon emission in the initial or final state]
are of the order of the electromagnetic coupling constant and should lead to
corrections at the level of a few per cent only. These corrections can thus
be safely neglected in a first stage.

\subsubsection*{3.3 Loop-induced decays}

\nn {\bf a) Radiative decays of the next-to-lightest neutralino (and gluino)}\s

If the mass splitting between the next-to-lightest neutralino and the lightest
neutralino is very small, the two-body neutralino decays discussed previously 
are not allowed kinematically; the virtuality of the exchanged particles
in the possible three-body decay modes [to be discussed later] is so large that
the loop-induced decays of at least $\chi_2^0$ into the LSP $\chi_1^0$ and a 
photon \cite{loop2,loop3}
\beq
\chi_2^0 \to \chi_1^0 \gamma 
\eeq
might be relevant. This decay is induced by triangle diagrams involving the
contribution of virtual charginos together with $W$ and charged Higgs bosons, 
and contributions with charged sfermion/fermion loops. It is of  ${\cal O}
(\alpha^3)$ in the electroweak coupling, but can have a sizeable branching
ratio [with respect to the three-body decays] in some areas of the MSSM 
parameter space and in some corners of the phase space. \s

For the full analytical formulae of the decay amplitudes, we use the ones that
were given in Ref.~\cite{loop2}, which are closer to our Lagrangian
convention and leads to an easier implementation in our program.  All diagrams
and contributions have been taken into account.\s

For completeness, and although this mode is never very important in the MSSM,
we have extended the previous calculation to the case where the gluino decays 
into a gluon and the lightest neutralino \cite{gluinorad}
\beq
\tilde{g} \to g \chi_1^0 \ ,
\eeq
which is mediated by only a subset of the diagrams involved in the previous
decay mode [that is, only the squark-quark loop contributions].\bigskip

\nn {\bf b) Loop-induced decay of the lightest top squark}\s

The heaviness of the top quark leads to distinct phenomenological  features for
the decays of its scalar partners. Indeed, while the other squarks can decay
directly into (almost) massless quarks and the lightest neutralino $\chi_1^0$,
which is always kinematically accessible,  since in general, in the MSSM the
neutralino $\chi_1^0$ is assumed to be the LSP, the decay channels $\tilde{t}_i
\to t \chi_1^0$  are kinematically closed for $ m_{\tilde{t}_i} \leq m_t
+m_{\chi_1^0}$.  If, in addition, $ m_{\tilde{t}_i} \leq m_b +m_{\chi_1^+}$,
the decay mode into a chargino and a $b$ quark, $\tilde{t}_i \to b \chi_1^+$,
is not accessible and the only two-body decay channel that would be  allowed
is the loop-induced and FCNC decay \cite{loop1}:
\beq
\tilde{t}_i \to c \chi_1^0
\eeq
This mode is mediated by one-loop diagrams: vertex diagrams as well as squark
and quark self-energy diagrams [where bottom squarks, charginos, charged $W$
and Higgs bosons are running in the loops]. The flavour transition $b \ra c$
occurs through the charged currents. Adding the various contributions, a
divergence is left out, which must be subtracted by adding a counterterm to the
scalar self-mass diagrams. When working in mSUGRA-type models, where the squark
masses are unified at the GUT scale, the divergence is subtracted using a
soft counterterm at $\Lambda_{\rm GUT}$, generating a large residual logarithm
$\log(\Lambda^2_{\rm GUT}/M_W^2) \sim 65$ in the amplitude. This logarithm
gives the leading contribution to the $\tilde{t}_1 \ra c \chi_1^0$ amplitude
and makes the decay width rather large, although it is suppressed
by the CKM matrix element $V_{cb} \sim 0.05$ and the (running) $b$ quark mass
squared $m_b^2 \sim (3$ GeV$)^2$. It is this approach that we have implemented
in the program {\tt SDECAY} and we have used the approximate formulae of
Ref.~\cite{loop1} [the exact expressions of the loop amplitudes are not yet
available].  \s

However, there are scenarios in which the decay rate $\Gamma(\tilde{t}_1 \ra 
c \chi_1^0$) can be rather small: $(i)$ First, the large logarithm $\log
\left(\Lambda_{\rm GUT}^2/ M_W^2 \right) \sim 65$ appears only because the
choice of the renormalization condition is made at $\Lambda_{\rm GUT}$, but in
a general MSSM, where the squark masses are not unified at some very high scale,
one could chose a low-energy counterterm; in this case no large logarithm would
appear.  $(ii)$ If the lightest top squark is a pure right-handed state [as
favoured by the constraints from high-precision electroweak data],
the amplitude involves only one component, which can be made small by choosing
small values of the trilinear coupling $A_b$ and/or large values of the
(common) SUSY-breaking scalar mass $\tilde{m}_q$. $(iii)$ Even in the presence
of stop mixing, for a given choice of MSSM parameters, large cancellations 
can occur between the various terms in the loop amplitude; in addition, the
$t$--$\tilde{t}_1$--$\chi_1^0$ coupling, which enters as a global factor, can 
be very tiny. \s

Thus, the decay rate $\Gamma (\tilde{t}_1 \ra c \chi_1^0)$ might be very small,
opening the possibility for the three-body and even four-body decay modes,
which will be discussed now,  to dominate. 

\subsubsection*{3.4 Multibody decay modes}

If the tree-level two-body decay modes discussed previously are kinematically
closed, i.e. equal to zero, multibody final state channels 
[as well as the loop decays
which have been discussed previously] will be the dominant decays and should be
considered.  There are three varieties of such decays with which the program
{\tt SDECAY}  deals, and which will be discussed below.  In all cases, we make
a clear separation between the two-body and multibody decays: that is, in the
higher order decay modes, we do not include the total decay widths of the
virtually exchanged  (s)particles in their respective propagators, so as to
make smooth transitions between the two-body and the multibody  decay
modes\footnote{This aspect leads to very complicated technical problems and
numerical instabilities. Indeed, once the total decay widths of the virtual
particles are included, the phase-space integrals cannot be made analytically
and one has to resort to a numerical integration which gives less precise
results and heavily slows down the program. In addition, there are issues of
gauge invariance that are not yet settled in this case, and which we did not
want to address at the moment.}. \bigskip

\nn {\bf a) Three-body decays of charginos, neutralinos and gluinos}\s

When the two-body channels of the charginos and neutralinos given in
eqs.~(5--7) are kinematically closed, the particles can decay into three-body
final states involving a lighter ino and two massless fermions:
\beq 
\label{decay}
\chi_i \, \rightarrow \, \chi_j  f \bar{f} \ .
\eeq
In the past, these decays have been discussed in particular for the lightest 
chargino
and for the next-to-lightest neutralino, when decaying into the LSP neutralino
and two fermions, $\chi_1^+ \to \chi^0_1 f \bar{f'}$ and $\chi_2^0 \to \chi_1^0
f \bar{f}$. In our case, we will not assume that the initial inos are
$\chi_1^+$ and $\chi_1^0$ and the final neutralino is the LSP $\chi_1^0$, but
any of the charginos $\chi_i^+$ ($i=1,2$) and neutralinos $\chi_i^0$ 
($i=1,...,4$) to cover all situations and the possibility
of cascade decays. These decays proceed through gauge boson exchange [$V=W$
and $Z$ for $\chi_i^+$ and $\chi_i^0$ decays, respectively], Higgs boson
exchange [$\Phi=H^+$ for $\chi_i^+$ decays and $\Phi=H,h,A$ for $\chi_i^0$
decays] and sfermion exchange in the $t$- and $u$-channels [the flavour is
fixed  by the sfermion-fermion and final neutralino vertex].  \s

The gluino can also undergo three-body decays  when the two-body decays of
eq.~(8) are kinematically forbidden: 
\beq
\tilde{g} \to \chi q \bar{q} \ ,
\eeq
and only the channels with $t$- and $u$-channel squark exchange will be present
in this case; the partial widths can be straightforwardly derived from those of
the neutralino decays, with the appropriate change of the couplings and 
the QCD factors. We also implemented the decay
\beq
\tilde{g} \to \tilde{t}_1 \bar{b} W^-
\eeq
which can be important in regions of the parameter space where it is 
kinematically allowed \cite{porodgluino}. Furthermore the decay 
\cite{dattagluino} 
\beq
\tilde{g} \to \tilde{t}_1 \bar{b} H^-
\eeq
has been taken into account.
\s

In fact, the heavier charginos and neutralinos can also decay [in particular,
in models with non-universal gaugino masses at high scale] into gluinos
and quark-antiquark pairs
\beq
\chi_i \to \tilde{g} q\bar{q} \ ,
\eeq
and the amplitude can be adapted from the one of the reverse process 
eq.~(16) discussed above.\s

The program {\tt SDECAY} calculates the partial widths and branching ratios of
all these three-body decay channels, taking into account all the possible 
contributions of the virtual particles. The radiatively corrected Yukawa
couplings of third-generation fermions, the mixing pattern for their sfermion
partners and the masses of the sparticles and gauge/Higgs bosons involved in
the processes are taken into account. In fact, even the masses of the fermion 
final states have been taken into account, since finite fermion masses are
needed in some cases\footnote{The approximation of zero mass is rather good
for all light fermion final states, except for $b$-quarks and $\tau$-leptons
when  $\chi_2^0$ and $\chi_1^+$ have masses close to the $\chi_1^0$ mass; the
zero-mass approximation would be very bad for top quark final states.
Nevertheless, in mSUGRA-type models, if the three-body decays $\chi_2^0 \to
\chi_1^0 t \bar{t}$ and $\chi_1^+ \to \chi_1^0 t \bar{b}$ are kinematically
allowed, they will not play a major role, since the charginos and neutralinos
will have enough phase space to decay first into the two-body channels
$\chi_2^0 \to \chi_1^0 Z,\; \chi_1^0 h$ [and possibly $\chi_1^0 H$ and
$\chi_1^0 A$] and $\chi_1^+ \to  \chi_1^0 W$ [and possibly $\chi_1^0 H^+$],
which will be largely dominating.}. We have used the  analytical formulae  of
Ref.~\cite{3body2} for the  matrix elements squared and integrated numerically
over the three-body phase space. \bigskip

\nn {\bf b) Three-body decays of third-generation sfermions}\s

For relatively heavier top squarks, when their masses are larger than the mass
of the $\chi_1^0$ neutralino and the mass of the $W$ boson, the $H^\pm$ boson
and/or the sfermion  $\tilde{f}^*$, there will be the possibility of the
three-body decay modes of the $\tilde{t}_1$. These modes have been  discussed in
Refs.~\cite{3body3,3body4}.  When kinematically possible, some of these
channels can  dominate over the loop-induced  $\tilde{t}_i \to c \chi_1^0$
mode in rather large areas of the MSSM parameter space.\s

For $m_{\tilde{t}_1} > M_{W(H^\pm)}+m_{\chi_1^0}$,  the three-body decay 
channels into $W$ or charged Higgs bosons, a bottom quark and the LSP 
neutralino,  
\beq 
\tilde{t}_i \ \to \ b W^+ \chi_1^0 \ \ , \ \ b H^+ \chi_1^0
\eeq
can be accessible and  have been shown to be [at least for the one with $W$ 
boson final states] often  dominant in the case where $m_{\tilde{t}_1} \leq m_t
+m_{\chi_1^0}$ and $m_b + m_{\chi_1^+}$. In addition, if sleptons are lighter
than squarks [as is often  the case in models with a common scalar mass at the
GUT scale, such as the mSUGRA model] the modes
\beq 
\tilde{t}_i \ \to \ b  l^+ \tilde{\nu}_l\ \ {\rm and/or} \ \ b \tilde{l}^+ \nu_l
\eeq
become possible. In the case of the lightest top  squarks,
they can be largely dominating over the loop-induced $c\chi_1^0$ mode. \s

In fact, these three-body decay modes are important not only for the lightest
top squark, but also for the heavier one. In addition, there is another
possibility which is the decay of the top squarks into a fermion-antifermion
pair and the lightest $\tilde{b}$ state [which can become the lightest scalar
quark in the case where $\tb$ is very large],  which is mediated by the virtual
exchange of $W$ and $H^+$ bosons: 
\beq
\tilde{t}_i \ \to \ \tilde{b}_1 \, f \bar{f}' \ .
\eeq
For the heavier top squark $\tilde{t}_2$, another possibility would be the
three-body decay into the lightest top squark and a fermion pair [with $f
\neq b$] through the exchange of the $Z$ and the MSSM neutral Higgs bosons 
[the CP-even $h,H$ and the CP-odd $A$ bosons],
\beq
\tilde{t}_2 \ \to \ \tilde{t}_1 \, f \bar{f} \ . 
\eeq
These modes apply also for the charged decays of heavier bottom squarks
into top squarks (and vice versa) which, as previously, occur through 
$W$ and $H^+$ boson exchanges 
\beq
\tilde{q}_2 \ \to \ \tilde{q}_j' \, f \bar{f}' \ .
\eeq
[If the mass splitting between the  initial and final scalar eigenstates is
large enough, the gauge and Higgs  bosons become real, and we have the two-body
decays into gauge and Higgs  bosons which have been discussed previously.]\s

For $b\bar{b}$ final states, one needs to include, in the case of $\tilde{t}_2 
\to \ \tilde{t}_1 \, b \bar{b}$, the contributions of the exchange of the two
charginos states $\chi_{1,2}^+$. This is also the case of the decay mode
$\tilde{b}_2 \ \to \ \tilde{b}_1 \, b \bar{b}$, where one has, in addition, the
virtual exchange of neutralinos and gluinos, which have to be taken into
account. The latter process is a generalization [since the mixing pattern is
more complicated] of the decay modes of first- and second-generation squarks
into light scalar bottoms \cite{3body4}, and would be in competition with at
least the two-body mode $\tilde{b}_2 \to b \chi_1^0$. The latter channel is
always open since $\chi_1^0$ is the LSP, but the $b$--$\tilde{b}_2$--$\chi_1^0$
coupling can be small, leaving the possibility to the three-body mode to occur
at a sizeable rate.  \s

{\tt SDECAY} evaluates all the three-body decay modes of the top squarks 
discussed above, when  the corresponding two-body decay channels are 
kinematically closed. In all cases, 
the  analytical expressions for the Dalitz plot densities in terms of the
energies of the final fermions \cite{3body4}, which take account of all
contributing channels, mixing and masses [including non-zero third-generation
fermion masses], are integrated numerically. As in the case of the inos, the
total decay widths of the exchanged particles are not included in the
propagators of the virtual particles. The three-body decays of the 
bottom squarks will be included in the upgraded version of {\tt SDECAY}. 
\bigskip

\nn {\bf c) The four-body decay of the top squark}\s

If the previous three-body decay modes of the top squarks are kinematically not
accessible, the  main $\tilde{t}_1$ decay channel is then expected to be the
loop-induced and flavor-changing decay into a charm quark and the LSP,
$\tilde{t}_1 \ra c\chi_1^0$. However, there is another decay mode that is
possible in the MSSM, even if $\tilde{t}_1$ is the next-to-lightest SUSY
particle [provided that $m_{\tilde{t}_1} > m_{\chi_1^0}+m_b$]: the four-body
decay into a $b$-quark, the LSP and two massless fermions \cite{4body}:
\beq
\tilde{t}_1 \ra b\chi_1^0 f\bar{f}' \ .
\eeq
This mode is of the same order of perturbation theory as the decay $\tilde{t}_1
\ra c \chi_1^0$, i.e. ${\cal O} (\alpha^3)$; in principle, it can therefore
compete with the latter channel. \s

The four-body decay mode  proceeds  through several diagrams: there are first
the $W$-boson  exchange diagrams with virtual $t, \tilde{b}$ and
$\chi^\pm_{1,2}$  states, a similar set of diagrams is obtained by replacing
the $W$-boson  by the $H^+$-boson and a third type of diagrams consist of up
and down type  slepton and first/second-generation squark exchanges. The decay
rate has been calculated in Ref.~\cite{4body} taking into account all diagrams
and  interferences. The various contributions can be summarized as: \s

$i)$ Because in the MSSM, $H^\pm$ has a mass larger than $M_W$ and has tiny
Yukawa couplings to light fermions, it does not give rise to large 
contributions. The squark exchange diagrams give also small contributions 
since squarks are expected to be much heavier than the $\tilde{t}_1$ state. 
The contribution of the diagram with an exchanged $t$-quark is
only  important if the stop mass is of the order of $m_t+m_{\chi_1^0} \gsim
{\cal O} (250$ GeV). 

$ii)$ In contrast to squarks, slepton [and especially $\tilde{\nu}$] exchange 
diagrams might give substantial contributions, since $\tilde{l}$ masses of 
${\cal O}$(100 GeV) are still experimentally allowed \cite{PDG}. In fact, when 
the difference between $m_{\tilde{t}_1}$ and $m_{\chi_1^+}$ and
$m_{\tilde{l}}$  is not large, this diagram will give the dominant contribution
to the four-body decay mode.  

$iii)$ The most significant contributions to the four-body decay mode will 
come in general from the diagram in which the lightest chargino $\chi_1^+$ and
the $W$ boson are exchanged, when the virtuality of the chargino is not too
large. In particular, for an exchanged $\chi_1^+$ with a mass not much larger
than $100$ GeV, the decay width can  be substantial even for top squark masses
of the order of 100 GeV. \s

Thus, a good approximation [especially for a light top squark $m_{\tilde{t}_1}
\sim {\cal O}(100$ GeV)] is to take into account only the top quark, the
lightest chargino and the slepton exchange diagrams. All the other
contributions, in particular the $H^\pm$ contribution,  can be safely neglected
in general. This is the approach that we choose in {\tt SDECAY}: we include all
the contributions and the interference effects, except for the small $H^\pm$
contribution which is very lengthy and time consuming to evaluate. The
four-body phase-space integrals are evaluated using the  program {\tt Rambo}. 

\subsubsection*{3.5 Decays of the top quark}

The decays of the heavy top quark in the MSSM that have also been included in 
the  program consist of the standard decay into a $W$ boson and a bottom 
quark, 
\beq
t \to bW^+ 
\eeq
but also, if allowed by phase space, of the decays into a bottom quark and an
$H^\pm$ boson and into a lighter top squark and a neutralino
\beq
t \to bH^+ \ {\rm and} \ \tilde{t}_1 \chi_1^0
\eeq
In addition, the one-loop QCD corrections \cite{QCDtop} will be included in
the next version of the program.  

\subsection*{4. Running {\tt SDECAY}}

\subsubsection*{4.1 Basic facts about {\tt SDECAY}}

Besides the files of the program {\tt SuSpect}, i.e. the input file {\tt
suspect2.in} [where one can select the model to be investigated, the 
accuracy of
the  spectrum algorithm, the input data (SM fermion masses and gauge couplings)
as well as SUSY and soft SUSY-breaking parameters] and the main Fortran routine
{\tt suspect2.f} [where some internally documented changes have been performed
for a fully consistent calculation, and the calling routine {\tt
suspect2\_call.f} is not needed], where the calculation of the spectrum is
performed and which needs the routines {\tt subh\_hdec.f}, {\tt feynhiggs.f},
{\tt hmsusy.f} and {\tt Hmasses} for the calculation of the Higgs boson
masses,  the program {\tt SDECAY} is composed of three files:\s

1) \underline{The input file {\tt sdecay.in}}: in this file,  one can choose 
the accuracy of the algorithm and make the choice of the various options:
whether QCD corrections and multibody or loop decays are included, which scales
are used for the couplings, the number of loops in their running and if top and
GMSB decays are to be evaluated  or not.\s 

2) \underline{The main routine {\tt sdecay.f}} where the couplings of the SUSY
particles and Higgs bosons are evaluated and the decay branching ratios and
total widths are calculated. This routine is self-contained and includes all
the necessary files for the calculation. \s

3) \underline{The output file {\tt sdecay.out}}: this file is  generated  at
each run of the program and gives the results for the  output branching ratios
and total decay widths. Two formats are possible for this file: either the
Higgs and SUSY particles are denoted in a simple and transparent form, or the
PDG notation is used. In addition, the masses of the SUSY and Higgs particles,
the mixing matrices and the gauge and third-generation couplings at the EWSB or
a chosen scale are given. [An additional output {\tt suspect2.out} is also
generated for the spectrum and also includes the soft SUSY-breaking terms].\s 

The routine {\tt sdecay.f} consists of about 30.000 lines of code and takes 
about 1 Mo of memory, while the input file only has a  few dozen lines (most of
them comments). The accompanying  routines for the calculation of the sparticle
and Higgs masses, which are provided separately,  have  in total a comparable
size.  The complete executable file takes about 2.5 Mo of disk space. The
running time for a  typical model point, for instance the mSUGRA point
discussed below, is a few seconds on a PC with a 1 GHz processor.\s

The Fortran files have to be compiled altogether and, running for
instance  on a PC, the compilation and link commands are [they are provided 
in a makefile]:
\begin{verbatim}
  OBJS = suspect2.o subh_hdec.o feynhiggs.o hmsusy.o sdecay.o
  FC=f77
  .f.o: 
        $(FC) -c $*.f
  sdecay: $(OBJS)
        $(FC) $(OBJS) -o run
\end{verbatim}

Thus, the program {\tt SDECAY} has the following structure:

\begin{itemize}
\vspace*{-2mm}

\item[--] It reads all the inputs in the files {\tt suspect2.in} and {\tt
sdecay.in}. \vspace*{-2mm}

\item[--] It calculates the sparticle and Higgs boson masses as well as
all the soft SUSY-breaking parameters using the program {\tt SuSpect}.
\vspace*{-2mm}

\item[--] It calls the subroutine {\tt common\_ini} where all parameters 
necessary  for the calculation of the couplings and decay widths are set.
\vspace*{-2mm}

\item[--] It calls the subroutine {\tt couplings} where all couplings necessary
for the calculation of the decay widths are evaluated.
\vspace*{-2mm}

\item[--] It calls the subroutines for the two-body decays (2), the three-body
decays (3)  and loop decays (l) and for the stop four-body decay (4) 
calculation of the total widths and branching ratios of the respective 
decaying particle:
\begin{verbatim}
chargino decays:      subroutines: char2bod (2), xintegchipm (3)
neutralino decays:    subroutines: neut2bod (2), xintegneut (3), 
                                   neutraddecay (l)
gluino decays:        subroutines: glui2bod (2), xinteggo (3),
                                   gluiraddecay (l)
sup decays:           subroutine : sup2bod (2)
sdown decays:         subroutine : sdown2bod (2)
stop decays:          subroutines: st2bod (2), xintegstop (3), 
                                   hikasakob1 (l), st4bod (4)
sbottom decays:       subroutine : sb2bod (2) 
selectron decays:     subroutine : sel2bod (2)
sneutrino_el decays:  subroutine : snel2bod (2)
stau decays:          subroutine : stau2bod (2)
sneutrino_tau decays: subroutine : sntau2bod (2)
\end{verbatim}  
The routines call several help functions and subroutines for the loop
decays, the QCD corrections, and some matrix elements for the multibody 
decays. They can be found at the end of the program with some comments 
specifying their purposes and their main features.
\vspace*{-2mm}

\item[--] It writes in the output file {\tt sdecay.out} where you can choose
two  versions: the output \`a la Les Houches Accord or one that is
easier  to read. \vspace*{-2mm} 
\end{itemize}

In the next subsections, we will exhibit the input and output files, taking 
the example of an mSUGRA benchmark point from the Snowmass Points and Slopes 
\cite{benchmark}, the so-called SPS1a point, with the inputs at the high scale:
\beq
m_0=100~{\rm GeV}\ , \ m_{1/2}=250~{\rm GeV}
\ , \ A_0=-100 \ , \  \tb=10 \ , \ \mu>0 \ . \non 
\eeq
The input and output files are self-explanatory and will not be 
commented  further.

\subsubsection*{4.2 The input file}

\begin{verbatim}
                          SDECAY INPUT FILE
                          -----------------

* Choice of the output, Les Houches Accord (1) or simple (0):
1

* Include (1) or not (0) the QCD corrections to the 2-body decay widths:
0

* Include (1) or not (0) the multi-body decays for inos and stops:
1

* Include (1) or not (0) the loop induced decays for the gluino, 
  the neutralinos and stop1:
1

* Include (1) or not (0) the SUSY decays of the top quark:
1

* Include (1) or not (0) the possible decays of the NLSP in GMSB models:
  (ichoice(1) has to be set 11 in suspect2.in.)
0

* Scheme in which the running alphas and quark masses are calculated 
  if the scale is not the scale of EWSB: 
  (If QCD corrections are included, the DR_bar scheme has to be used.) 
  0 (MS_bar scheme) and 1 (DR_bar scheme).
1

* Scale at which the scale dependent couplings are calculated: 
  1: EWSB scale, 2: mass of the decaying sparticle, 3: user choice
1

* Scale of the scale dependent couplings if chosen by the user 
  (in GeV):
100.D0

* Number of loops for the calculation of the running couplings
2
\end{verbatim}

\subsubsection*{4.3 Output file according to SLHA}

\begin{verbatim}
# SUSY Les Houches Accord - MSSM Spectrum + Decays
# SDECAY 1.0
# Authors: M.Muhlleitner, A.Djouadi and Y.Mambrini
# In case of problems please send an email to
# margarete.muehlleitner@psi.ch
# djouadi@lpm.univ-montp2.fr
# mambrini@delta.ft.uam.es
#
# If not stated otherwise all couplings and masses
# are given at the scale of the electroweak symmetry
# breaking Q=  0.46296529E+03
#
#
BLOCK MASS  # Mass Spectrum
# PDG code           mass       particle
        25     1.14365068E+02   # h
        35     3.91956602E+02   # H
        36     3.92191912E+02   # A
        37     4.00353329E+02   # H+
   1000001     5.69828109E+02   # ~d_L
   2000001     5.43157826E+02   # ~d_R
   1000002     5.64244153E+02   # ~u_L
   2000002     5.44093303E+02   # ~u_R
   1000003     5.69828109E+02   # ~s_L
   2000003     5.43157826E+02   # ~s_R
   1000004     5.64244153E+02   # ~c_L
   2000004     5.44093303E+02   # ~c_R
   1000005     5.16713072E+02   # ~b_1
   2000005     5.44166483E+02   # ~b_2
   1000006     4.00256829E+02   # ~t_1
   2000006     5.80537860E+02   # ~t_2
   1000011     2.03724637E+02   # ~e_L
   2000011     1.45386789E+02   # ~e_R
   1000012     1.87810224E+02   # ~nu_eL
   1000013     2.03724637E+02   # ~mu_L
   2000013     1.45386789E+02   # ~mu_R
   1000014     1.87810224E+02   # ~nu_muL
   1000015     1.36395255E+02   # ~tau_1
   2000015     2.07528018E+02   # ~tau_2
   1000016     1.86942145E+02   # ~nu_tauL
   1000021     6.03561040E+02   # ~g
   1000022     9.89200644E+01   # ~chi_10
   1000023     1.76248916E+02   # ~chi_20
   1000025    -3.57870532E+02   # ~chi_30
   1000035     3.77017717E+02   # ~chi_40
   1000024     1.75568747E+02   # ~chi_1+
   1000037     3.77194407E+02   # ~chi_2+
#
BLOCK NMIX  # Neutralino Mixing Matrix
  1  1     9.84337446E-01   # N_11
  1  2    -6.25292026E-02   # N_12
  1  3     1.54406991E-01   # N_13
  1  4    -5.76920450E-02   # N_14
  2  1     1.12653886E-01   # N_21
  2  2     9.40622674E-01   # N_22
  2  3    -2.77623463E-01   # N_23
  2  4     1.59572240E-01   # N_24
  3  1     6.15283938E-02   # N_31
  3  2    -9.19478832E-02   # N_32
  3  3    -6.94945017E-01   # N_33
  3  4    -7.10500716E-01   # N_34
  4  1     1.20843498E-01   # N_41
  4  2    -3.20725226E-01   # N_42
  4  3    -6.45085357E-01   # N_43
  4  4     6.82932691E-01   # N_44
#
BLOCK UMIX  # Chargino Mixing Matrix U
  1  1    -9.12748331E-01   # U_11
  1  2     4.08522319E-01   # U_12
  2  1     4.08522319E-01   # U_21
  2  2     9.12748331E-01   # U_22
#
BLOCK VMIX  # Chargino Mixing Matrix V
  1  1    -9.71766502E-01   # V_11
  1  2     2.35944624E-01   # V_12
  2  1     2.35944624E-01   # V_21
  2  2     9.71766502E-01   # V_22
#
BLOCK STOPMIX  # Stop Mixing Matrix
  1  1     5.50903293E-01   # cos(theta_t)
  1  2     8.34569087E-01   # sin(theta_t)
  2  1    -8.34569087E-01   # -sin(theta_t)
  2  2     5.50903293E-01   # cos(theta_t)
#
BLOCK SBOTMIX  # Sbottom Mixing Matrix
  1  1     9.21378487E-01   # cos(theta_b)
  1  2     3.88666546E-01   # sin(theta_b)
  2  1    -3.88666546E-01   # -sin(theta_b)
  2  2     9.21378487E-01   # cos(theta_b)
#
BLOCK STAUMIX  # Stau Mixing Matrix
  1  1     2.80106527E-01   # cos(theta_tau)
  1  2     9.59968923E-01   # sin(theta_tau)
  2  1    -9.59968923E-01   # -sin(theta_tau)
  2  2     2.80106527E-01   # cos(theta_tau)
#
BLOCK ALPHA  # Higgs mixing
          -1.13249720E-01   # Mixing angle in the neutral Higgs boson sector
#
BLOCK HMIX Q=  4.62965294E+02  # DRbar Higgs Mixing Parameters
     1     3.51486069E+02   # mu
#
BLOCK GAUGE Q=  4.62965294E+02  # The gauge couplings
     1     3.62163400E-01   # gprime(Q) DRbar
     2     6.46905504E-01   # g(Q) DRbar
     3     1.09847635E+00   # g3(Q) DRbar
#
BLOCK AU, AD, AE Q=  4.62965294E+02  # The trilinear couplings
  3  3    -5.11225438E+02   # A_t DRbar
  3  3    -7.92584896E+02   # A_b DRbar
  3  3    -2.54143182E+02   # A_tau DRbar
#
BLOCK Y_X,A_X Q=  4.62965294E+02  # The Yukawa couplings
  3  3     1.35709601E+00   # y_t DRbar
  3  3     2.10275093E-01   # y_b DRbar
  3  3     1.55612239E-01   # y_tau DRbar
#
#
#
# =================
# |The decay table|
# =================
#
# The multi-body decays for the inos and sfermions are included.
#
# The loop induced decays for the gluino, neutralinos and stops
# are included.
#
# The SUSY decays of the top quark are included.
#
#
#         PDG            Width
DECAY         6     1.50609870E+00   # top decays
#          BR         NDA      ID1       ID2
     1.00000000E+00    2           5        24   # BR(t ->  b    W+)
     0.00000000E+00    2           5        37   # BR(t ->  b    H+)
     0.00000000E+00    2     1000006   1000022   # BR(t -> ~t_1 ~chi_10)
     0.00000000E+00    2     1000006   1000023   # BR(t -> ~t_1 ~chi_20)
     0.00000000E+00    2     1000006   1000025   # BR(t -> ~t_1 ~chi_30)
     0.00000000E+00    2     1000006   1000035   # BR(t -> ~t_1 ~chi_40)
     0.00000000E+00    2     2000006   1000022   # BR(t -> ~t_2 ~chi_10)
     0.00000000E+00    2     2000006   1000023   # BR(t -> ~t_2 ~chi_20)
     0.00000000E+00    2     2000006   1000025   # BR(t -> ~t_2 ~chi_30)
     0.00000000E+00    2     2000006   1000035   # BR(t -> ~t_2 ~chi_40)
#
#         PDG            Width
DECAY   1000021     4.85459975E+00   # gluino decays
#          BR         NDA      ID1       ID2
     1.76180098E-02    2     1000001        -1   # BR(~g -> ~d_L  db)
     1.76180098E-02    2    -1000001         1   # BR(~g -> ~d_L* d )
     5.39508838E-02    2     2000001        -1   # BR(~g -> ~d_R  db)
     5.39508838E-02    2    -2000001         1   # BR(~g -> ~d_R* d )
     2.37062902E-02    2     1000002        -2   # BR(~g -> ~u_L  ub)
     2.37062902E-02    2    -1000002         2   # BR(~g -> ~u_L* u )
     5.23780813E-02    2     2000002        -2   # BR(~g -> ~u_R  ub)
     5.23780813E-02    2    -2000002         2   # BR(~g -> ~u_R* u )
     1.76180098E-02    2     1000003        -3   # BR(~g -> ~s_L  sb)
     1.76180098E-02    2    -1000003         3   # BR(~g -> ~s_L* s )
     5.39508838E-02    2     2000003        -3   # BR(~g -> ~s_R  sb)
     5.39508838E-02    2    -2000003         3   # BR(~g -> ~s_R* s )
     2.37062902E-02    2     1000004        -4   # BR(~g -> ~c_L  cb)
     2.37062902E-02    2    -1000004         4   # BR(~g -> ~c_L* c )
     5.23780813E-02    2     2000004        -4   # BR(~g -> ~c_R  cb)
     5.23780813E-02    2    -2000004         4   # BR(~g -> ~c_R* c )
     1.01674937E-01    2     1000005        -5   # BR(~g -> ~b_1  bb)
     1.01674937E-01    2    -1000005         5   # BR(~g -> ~b_1* b )
     5.53319969E-02    2     2000005        -5   # BR(~g -> ~b_2  bb)
     5.53319969E-02    2    -2000005         5   # BR(~g -> ~b_2* b )
     4.76865361E-02    2     1000006        -6   # BR(~g -> ~t_1  tb)
     4.76865361E-02    2    -1000006         6   # BR(~g -> ~t_1* t )
     0.00000000E+00    2     2000006        -6   # BR(~g -> ~t_2  tb)
     0.00000000E+00    2    -2000006         6   # BR(~g -> ~t_2* t )
#
#         PDG            Width
DECAY   1000006     2.06530974E+00   # stop1 decays
#          BR         NDA      ID1       ID2
     1.76720929E-01    2     1000022         6   # BR(~t_1 -> ~chi_10 t )
     1.26493082E-01    2     1000023         6   # BR(~t_1 -> ~chi_20 t )
     0.00000000E+00    2     1000025         6   # BR(~t_1 -> ~chi_30 t )
     0.00000000E+00    2     1000035         6   # BR(~t_1 -> ~chi_40 t )
     6.78820856E-01    2     1000024         5   # BR(~t_1 -> ~chi_1+ b )
     1.79651325E-02    2     1000037         5   # BR(~t_1 -> ~chi_2+ b )
     0.00000000E+00    2     1000021         6   # BR(~t_1 -> ~g      t )
     0.00000000E+00    2     1000005        37   # BR(~t_1 -> ~b_1    H+)
     0.00000000E+00    2     2000005        37   # BR(~t_1 -> ~b_2    H+)
     0.00000000E+00    2     1000005        24   # BR(~t_1 -> ~b_1    W+)
     0.00000000E+00    2     2000005        24   # BR(~t_1 -> ~b_2    W+)
#
#         PDG            Width
DECAY   2000006     7.07039620E+00   # stop2 decays
#          BR         NDA      ID1       ID2
     2.63043489E-02    2     1000022         6   # BR(~t_2 -> ~chi_10 t )
     9.04476605E-02    2     1000023         6   # BR(~t_2 -> ~chi_20 t )
     4.42222844E-02    2     1000025         6   # BR(~t_2 -> ~chi_30 t )
     1.96823757E-01    2     1000035         6   # BR(~t_2 -> ~chi_40 t )
     2.29818044E-01    2     1000024         5   # BR(~t_2 -> ~chi_1+ b )
     2.01693594E-01    2     1000037         5   # BR(~t_2 -> ~chi_2+ b )
     0.00000000E+00    2     1000021         6   # BR(~t_2 -> ~g      t )
     3.81187184E-02    2     1000006        25   # BR(~t_2 -> ~t_1    h )
     0.00000000E+00    2     1000006        35   # BR(~t_2 -> ~t_1    H )
     0.00000000E+00    2     1000006        36   # BR(~t_2 -> ~t_1    A )
     0.00000000E+00    2     1000005        37   # BR(~t_2 -> ~b_1    H+)
     0.00000000E+00    2     2000005        37   # BR(~t_2 -> ~b_2    H+)
     1.72571593E-01    2     1000006        23   # BR(~t_2 -> ~t_1    Z )
     0.00000000E+00    2     1000005        24   # BR(~t_2 -> ~b_1    W+)
     0.00000000E+00    2     2000005        24   # BR(~t_2 -> ~b_2    W+)
#
#         PDG            Width
DECAY   1000005     3.79216587E+00   # sbottom1 decays
#          BR         NDA      ID1       ID2
     4.61784398E-02    2     1000022         5   # BR(~b_1 -> ~chi_10 b )
     3.44388596E-01    2     1000023         5   # BR(~b_1 -> ~chi_20 b )
     5.08897163E-03    2     1000025         5   # BR(~b_1 -> ~chi_30 b )
     1.06407337E-02    2     1000035         5   # BR(~b_1 -> ~chi_40 b )
     4.49199440E-01    2    -1000024         6   # BR(~b_1 -> ~chi_1- t )
     0.00000000E+00    2    -1000037         6   # BR(~b_1 -> ~chi_2- t )
     0.00000000E+00    2     1000021         5   # BR(~b_1 -> ~g      b )
     0.00000000E+00    2     1000006       -37   # BR(~b_1 -> ~t_1    H-)
     0.00000000E+00    2     2000006       -37   # BR(~b_1 -> ~t_2    H-)
     1.44503819E-01    2     1000006       -24   # BR(~b_1 -> ~t_1    W-)
     0.00000000E+00    2     2000006       -24   # BR(~b_1 -> ~t_2    W-)
#
#         PDG            Width
DECAY   2000005     9.30237022E-01   # sbottom2 decays
#          BR         NDA      ID1       ID2
     2.18042454E-01    2     1000022         5   # BR(~b_2 -> ~chi_10 b )
     1.63625100E-01    2     1000023         5   # BR(~b_2 -> ~chi_20 b )
     4.87739508E-02    2     1000025         5   # BR(~b_2 -> ~chi_30 b )
     7.28430988E-02    2     1000035         5   # BR(~b_2 -> ~chi_40 b )
     2.17719337E-01    2    -1000024         6   # BR(~b_2 -> ~chi_1- t )
     0.00000000E+00    2    -1000037         6   # BR(~b_2 -> ~chi_2- t )
     0.00000000E+00    2     1000021         5   # BR(~b_2 -> ~g      b )
     0.00000000E+00    2     1000005        25   # BR(~b_2 -> ~b_1    h )
     0.00000000E+00    2     1000005        35   # BR(~b_2 -> ~b_1    H )
     0.00000000E+00    2     1000005        36   # BR(~b_2 -> ~b_1    A )
     0.00000000E+00    2     1000006       -37   # BR(~b_2 -> ~t_1    H-)
     0.00000000E+00    2     2000006       -37   # BR(~b_2 -> ~t_2    H-)
     0.00000000E+00    2     1000005        23   # BR(~b_2 -> ~b_1    Z )
     2.78996060E-01    2     1000006       -24   # BR(~b_2 -> ~t_1    W-)
     0.00000000E+00    2     2000006       -24   # BR(~b_2 -> ~t_2    W-)
#
#         PDG            Width
DECAY   1000002     5.56209623E+00   # sup_L decays
#          BR         NDA      ID1       ID2
     5.04724246E-03    2     1000022         2   # BR(~u_L -> ~chi_10 u)
     3.17395881E-01    2     1000023         2   # BR(~u_L -> ~chi_20 u)
     9.89628847E-04    2     1000025         2   # BR(~u_L -> ~chi_30 u)
     1.15831154E-02    2     1000035         2   # BR(~u_L -> ~chi_40 u)
     6.50599773E-01    2     1000024         1   # BR(~u_L -> ~chi_1+ d)
     1.43843593E-02    2     1000037         1   # BR(~u_L -> ~chi_2+ d)
     0.00000000E+00    2     1000021         2   # BR(~u_L -> ~g      u)
#
#         PDG            Width
DECAY   2000002     1.05910452E+00   # sup_R decays
#          BR         NDA      ID1       ID2
     9.83357095E-01    2     1000022         2   # BR(~u_R -> ~chi_10 u)
     1.10363000E-02    2     1000023         2   # BR(~u_R -> ~chi_20 u)
     1.32287897E-03    2     1000025         2   # BR(~u_R -> ~chi_30 u)
     4.28372575E-03    2     1000035         2   # BR(~u_R -> ~chi_40 u)
     0.00000000E+00    2     1000024         1   # BR(~u_R -> ~chi_1+ d)
     0.00000000E+00    2     1000037         1   # BR(~u_R -> ~chi_2+ d)
     0.00000000E+00    2     1000021         2   # BR(~u_R -> ~g      u)
#
#         PDG            Width
DECAY   1000001     5.35470938E+00   # sdown_L decays
#          BR         NDA      ID1       ID2
     2.35733116E-02    2     1000022         1   # BR(~d_L -> ~chi_10 d)
     3.07006812E-01    2     1000023         1   # BR(~d_L -> ~chi_20 d)
     1.72033892E-03    2     1000025         1   # BR(~d_L -> ~chi_30 d)
     1.64027859E-02    2     1000035         1   # BR(~d_L -> ~chi_40 d)
     6.04624253E-01    2    -1000024         2   # BR(~d_L -> ~chi_1- u)
     4.66724989E-02    2    -1000037         2   # BR(~d_L -> ~chi_2- u)
     0.00000000E+00    2     1000021         1   # BR(~d_L -> ~g      d)
#
#         PDG            Width
DECAY   2000001     2.64248238E-01   # sdown_R decays
#          BR         NDA      ID1       ID2
     9.83395634E-01    2     1000022         1   # BR(~d_R -> ~chi_10 d)
     1.10304127E-02    2     1000023         1   # BR(~d_R -> ~chi_20 d)
     1.31629502E-03    2     1000025         1   # BR(~d_R -> ~chi_30 d)
     4.25765842E-03    2     1000035         1   # BR(~d_R -> ~chi_40 d)
     0.00000000E+00    2    -1000024         2   # BR(~d_R -> ~chi_1- u)
     0.00000000E+00    2    -1000037         2   # BR(~d_R -> ~chi_2- u)
     0.00000000E+00    2     1000021         1   # BR(~d_R -> ~g      d)
#
#         PDG            Width
DECAY   1000004     5.56209623E+00   # scharm_L decays
#          BR         NDA      ID1       ID2
     5.04724246E-03    2     1000022         4   # BR(~c_L -> ~chi_10 c)
     3.17395881E-01    2     1000023         4   # BR(~c_L -> ~chi_20 c)
     9.89628847E-04    2     1000025         4   # BR(~c_L -> ~chi_30 c)
     1.15831154E-02    2     1000035         4   # BR(~c_L -> ~chi_40 c)
     6.50599773E-01    2     1000024         3   # BR(~c_L -> ~chi_1+ s)
     1.43843593E-02    2     1000037         3   # BR(~c_L -> ~chi_2+ s)
     0.00000000E+00    2     1000021         4   # BR(~c_L -> ~g      c)
#
#         PDG            Width
DECAY   2000004     1.05910452E+00   # scharm_R decays
#          BR         NDA      ID1       ID2
     9.83357095E-01    2     1000022         4   # BR(~c_R -> ~chi_10 c)
     1.10363000E-02    2     1000023         4   # BR(~c_R -> ~chi_20 c)
     1.32287897E-03    2     1000025         4   # BR(~c_R -> ~chi_30 c)
     4.28372575E-03    2     1000035         4   # BR(~c_R -> ~chi_40 c)
     0.00000000E+00    2     1000024         3   # BR(~c_R -> ~chi_1+ s)
     0.00000000E+00    2     1000037         3   # BR(~c_R -> ~chi_2+ s)
     0.00000000E+00    2     1000021         4   # BR(~c_R -> ~g      c)
#
#         PDG            Width
DECAY   1000003     5.35470938E+00   # sstrange_L decays
#          BR         NDA      ID1       ID2
     2.35733116E-02    2     1000022         3   # BR(~s_L -> ~chi_10 s)
     3.07006812E-01    2     1000023         3   # BR(~s_L -> ~chi_20 s)
     1.72033892E-03    2     1000025         3   # BR(~s_L -> ~chi_30 s)
     1.64027859E-02    2     1000035         3   # BR(~s_L -> ~chi_40 s)
     6.04624253E-01    2    -1000024         4   # BR(~s_L -> ~chi_1- c)
     4.66724989E-02    2    -1000037         4   # BR(~s_L -> ~chi_2- c)
     0.00000000E+00    2     1000021         3   # BR(~s_L -> ~g      s)
#
#         PDG            Width
DECAY   2000003     2.64248238E-01   # sstrange_R decays
#          BR         NDA      ID1       ID2
     9.83395634E-01    2     1000022         3   # BR(~s_R -> ~chi_10 s)
     1.10304127E-02    2     1000023         3   # BR(~s_R -> ~chi_20 s)
     1.31629502E-03    2     1000025         3   # BR(~s_R -> ~chi_30 s)
     4.25765842E-03    2     1000035         3   # BR(~s_R -> ~chi_40 s)
     0.00000000E+00    2    -1000024         4   # BR(~s_R -> ~chi_1- c)
     0.00000000E+00    2    -1000037         4   # BR(~s_R -> ~chi_2- c)
     0.00000000E+00    2     1000021         3   # BR(~s_R -> ~g      s)
#
#         PDG            Width
DECAY   1000011     2.53683595E-01   # selectron_L decays
#          BR         NDA      ID1       ID2
     4.19337361E-01    2     1000022        11   # BR(~e_L -> ~chi_10 e-)
     2.11871941E-01    2     1000023        11   # BR(~e_L -> ~chi_20 e-)
     0.00000000E+00    2     1000025        11   # BR(~e_L -> ~chi_30 e-)
     0.00000000E+00    2     1000035        11   # BR(~e_L -> ~chi_40 e-)
     3.68790698E-01    2    -1000024        12   # BR(~e_L -> ~chi_1- nu_e)
     0.00000000E+00    2    -1000037        12   # BR(~e_L -> ~chi_2- nu_e)
#
#         PDG            Width
DECAY   2000011     1.93168017E-01   # selectron_R decays
#          BR         NDA      ID1       ID2
     1.00000000E+00    2     1000022        11   # BR(~e_R -> ~chi_10 e-)
     0.00000000E+00    2     1000023        11   # BR(~e_R -> ~chi_20 e-)
     0.00000000E+00    2     1000025        11   # BR(~e_R -> ~chi_30 e-)
     0.00000000E+00    2     1000035        11   # BR(~e_R -> ~chi_40 e-)
     0.00000000E+00    2    -1000024        12   # BR(~e_R -> ~chi_1- nu_e)
     0.00000000E+00    2    -1000037        12   # BR(~e_R -> ~chi_2- nu_e)
#
#         PDG            Width
DECAY   1000013     2.53683595E-01   # smuon_L decays
#          BR         NDA      ID1       ID2
     4.19337361E-01    2     1000022        13   # BR(~mu_L -> ~chi_10 mu-)
     2.11871941E-01    2     1000023        13   # BR(~mu_L -> ~chi_20 mu-)
     0.00000000E+00    2     1000025        13   # BR(~mu_L -> ~chi_30 mu-)
     0.00000000E+00    2     1000035        13   # BR(~mu_L -> ~chi_40 mu-)
     3.68790698E-01    2    -1000024        14   # BR(~mu_L -> ~chi_1- nu_mu)
     0.00000000E+00    2    -1000037        14   # BR(~mu_L -> ~chi_2- nu_mu)
#
#         PDG            Width
DECAY   2000013     1.93168017E-01   # smuon_R decays
#          BR         NDA      ID1       ID2
     1.00000000E+00    2     1000022        13   # BR(~mu_R -> ~chi_10 mu-)
     0.00000000E+00    2     1000023        13   # BR(~mu_R -> ~chi_20 mu-)
     0.00000000E+00    2     1000025        13   # BR(~mu_R -> ~chi_30 mu-)
     0.00000000E+00    2     1000035        13   # BR(~mu_R -> ~chi_40 mu-)
     0.00000000E+00    2    -1000024        14   # BR(~mu_R -> ~chi_1- nu_mu)
     0.00000000E+00    2    -1000037        14   # BR(~mu_R -> ~chi_2- nu_mu)
#
#         PDG            Width
DECAY   1000015     1.34568207E-01   # stau_1 decays
#          BR         NDA      ID1       ID2
     1.00000000E+00    2     1000022        15   # BR(~tau_1 -> ~chi_10  tau-)
     0.00000000E+00    2     1000023        15   # BR(~tau_1 -> ~chi_20  tau-)
     0.00000000E+00    2     1000025        15   # BR(~tau_1 -> ~chi_30  tau-)
     0.00000000E+00    2     1000035        15   # BR(~tau_1 -> ~chi_40  tau-)
     0.00000000E+00    2    -1000024        16   # BR(~tau_1 -> ~chi_1-  nu_tau)
     0.00000000E+00    2    -1000037        16   # BR(~tau_1 -> ~chi_2-  nu_tau)
     0.00000000E+00    2     1000016       -37   # BR(~tau_1 -> ~nu_tauL H-)
     0.00000000E+00    2     1000016       -24   # BR(~tau_1 -> ~nu_tauL W-)
#
#         PDG            Width
DECAY   2000015     3.03123027E-01   # stau_2 decays
#          BR         NDA      ID1       ID2
     4.63906167E-01    2     1000022        15   # BR(~tau_2 -> ~chi_10  tau-)
     1.96659153E-01    2     1000023        15   # BR(~tau_2 -> ~chi_20  tau-)
     0.00000000E+00    2     1000025        15   # BR(~tau_2 -> ~chi_30  tau-)
     0.00000000E+00    2     1000035        15   # BR(~tau_2 -> ~chi_40  tau-)
     3.39434680E-01    2    -1000024        16   # BR(~tau_2 -> ~chi_1-  nu_tau)
     0.00000000E+00    2    -1000037        16   # BR(~tau_2 -> ~chi_2-  nu_tau)
     0.00000000E+00    2     1000016       -37   # BR(~tau_2 -> ~nu_tauL H-)
     0.00000000E+00    2     1000016       -24   # BR(~tau_2 -> ~nu_tauL W-)
     0.00000000E+00    2     1000015        25   # BR(~tau_2 -> ~tau_1   h)
     0.00000000E+00    2     1000015        35   # BR(~tau_2 -> ~tau_1   H)
     0.00000000E+00    2     1000015        36   # BR(~tau_2 -> ~tau_1   A)
     0.00000000E+00    2     1000015        23   # BR(~tau_2 -> ~tau_1   Z)
#
#         PDG            Width
DECAY   1000012     1.73484437E-01   # snu_eL decays
#          BR         NDA      ID1       ID2
     8.14895139E-01    2     1000022        12   # BR(~nu_eL -> ~chi_10 nu_e)
     4.97403776E-02    2     1000023        12   # BR(~nu_eL -> ~chi_20 nu_e)
     0.00000000E+00    2     1000025        12   # BR(~nu_eL -> ~chi_30 nu_e)
     0.00000000E+00    2     1000035        12   # BR(~nu_eL -> ~chi_40 nu_e)
     1.35364484E-01    2     1000024        11   # BR(~nu_eL -> ~chi_1+ e-)
     0.00000000E+00    2     1000037        11   # BR(~nu_eL -> ~chi_2+ e-)
#
#         PDG            Width
DECAY   1000014     1.73484437E-01   # snu_muL decays
#          BR         NDA      ID1       ID2
     8.14895139E-01    2     1000022        14   # BR(~nu_muL -> ~chi_10 nu_mu)
     4.97403776E-02    2     1000023        14   # BR(~nu_muL -> ~chi_20 nu_mu)
     0.00000000E+00    2     1000025        14   # BR(~nu_muL -> ~chi_30 nu_mu)
     0.00000000E+00    2     1000035        14   # BR(~nu_muL -> ~chi_40 nu_mu)
     1.35364484E-01    2     1000024        13   # BR(~nu_muL -> ~chi_1+ mu-)
     0.00000000E+00    2     1000037        13   # BR(~nu_muL -> ~chi_2+ mu-)
#
#         PDG            Width
DECAY   1000016     1.67039909E-01   # snu_tauL decays
#          BR         NDA      ID1       ID2
     8.36412025E-01    2     1000022        16   # BR(~nu_tauL -> ~chi_10 nu_tau)
     4.45980206E-02    2     1000023        16   # BR(~nu_tauL -> ~chi_20 nu_tau)
     0.00000000E+00    2     1000025        16   # BR(~nu_tauL -> ~chi_30 nu_tau)
     0.00000000E+00    2     1000035        16   # BR(~nu_tauL -> ~chi_40 nu_tau)
     1.18989955E-01    2     1000024        15   # BR(~nu_tauL -> ~chi_1+ tau-)
     0.00000000E+00    2     1000037        15   # BR(~nu_tauL -> ~chi_2+ tau-)
     0.00000000E+00    2     1000015       -37   # BR(~nu_tauL -> ~tau_1+ H-)
     0.00000000E+00    2     2000015       -37   # BR(~nu_tauL -> ~tau_2+ H-)
     0.00000000E+00    2     1000015       -24   # BR(~nu_tauL -> ~tau_1+ W-)
     0.00000000E+00    2     2000015       -24   # BR(~nu_tauL -> ~tau_2+ W-)
#
#         PDG            Width
DECAY   1000024     1.15215450E-02   # chargino1+ decays
#          BR         NDA      ID1       ID2
     0.00000000E+00    2     1000002        -1   # BR(~chi_1+ -> ~u_L     db)
     0.00000000E+00    2     2000002        -1   # BR(~chi_1+ -> ~u_R     db)
     0.00000000E+00    2    -1000001         2   # BR(~chi_1+ -> ~d_L*    u )
     0.00000000E+00    2    -2000001         2   # BR(~chi_1+ -> ~d_R*    u )
     0.00000000E+00    2     1000004        -3   # BR(~chi_1+ -> ~c_L     sb)
     0.00000000E+00    2     2000004        -3   # BR(~chi_1+ -> ~c_R     sb)
     0.00000000E+00    2    -1000003         4   # BR(~chi_1+ -> ~s_L*    c )
     0.00000000E+00    2    -2000003         4   # BR(~chi_1+ -> ~s_R*    c )
     0.00000000E+00    2     1000006        -5   # BR(~chi_1+ -> ~t_1     bb)
     0.00000000E+00    2     2000006        -5   # BR(~chi_1+ -> ~t_2     bb)
     0.00000000E+00    2    -1000005         6   # BR(~chi_1+ -> ~b_1*    t )
     0.00000000E+00    2    -2000005         6   # BR(~chi_1+ -> ~b_2*    t )
     0.00000000E+00    2     1000012       -11   # BR(~chi_1+ -> ~nu_eL   e+  )
     0.00000000E+00    2     1000014       -13   # BR(~chi_1+ -> ~nu_muL  mu+ )
     0.00000000E+00    2     1000016       -15   # BR(~chi_1+ -> ~nu_tau1 tau+)
     0.00000000E+00    2    -1000011        12   # BR(~chi_1+ -> ~e_L+    nu_e)
     0.00000000E+00    2    -2000011        12   # BR(~chi_1+ -> ~e_R+    nu_e)
     0.00000000E+00    2    -1000013        14   # BR(~chi_1+ -> ~mu_L+   nu_mu)
     0.00000000E+00    2    -2000013        14   # BR(~chi_1+ -> ~mu_R+   nu_mu)
     1.00000000E+00    2    -1000015        16   # BR(~chi_1+ -> ~tau_1+  nu_tau)
     0.00000000E+00    2    -2000015        16   # BR(~chi_1+ -> ~tau_2+  nu_tau)
     0.00000000E+00    2     1000022        24   # BR(~chi_1+ -> ~chi_10  W+)
     0.00000000E+00    2     1000023        24   # BR(~chi_1+ -> ~chi_20  W+)
     0.00000000E+00    2     1000025        24   # BR(~chi_1+ -> ~chi_30  W+)
     0.00000000E+00    2     1000035        24   # BR(~chi_1+ -> ~chi_40  W+)
     0.00000000E+00    2     1000022        37   # BR(~chi_1+ -> ~chi_10  H+)
     0.00000000E+00    2     1000023        37   # BR(~chi_1+ -> ~chi_20  H+)
     0.00000000E+00    2     1000025        37   # BR(~chi_1+ -> ~chi_30  H+)
     0.00000000E+00    2     1000035        37   # BR(~chi_1+ -> ~chi_40  H+)
#
#         PDG            Width
DECAY   1000037     2.51070848E+00   # chargino2+ decays
#          BR         NDA      ID1       ID2
     0.00000000E+00    2     1000002        -1   # BR(~chi_2+ -> ~u_L     db)
     0.00000000E+00    2     2000002        -1   # BR(~chi_2+ -> ~u_R     db)
     0.00000000E+00    2    -1000001         2   # BR(~chi_2+ -> ~d_L*    u )
     0.00000000E+00    2    -2000001         2   # BR(~chi_2+ -> ~d_R*    u )
     0.00000000E+00    2     1000004        -3   # BR(~chi_2+ -> ~c_L     sb)
     0.00000000E+00    2     2000004        -3   # BR(~chi_2+ -> ~c_R     sb)
     0.00000000E+00    2    -1000003         4   # BR(~chi_2+ -> ~s_L*    c )
     0.00000000E+00    2    -2000003         4   # BR(~chi_2+ -> ~s_R*    c )
     0.00000000E+00    2     1000006        -5   # BR(~chi_2+ -> ~t_1     bb)
     0.00000000E+00    2     2000006        -5   # BR(~chi_2+ -> ~t_2     bb)
     0.00000000E+00    2    -1000005         6   # BR(~chi_2+ -> ~b_1*    t )
     0.00000000E+00    2    -2000005         6   # BR(~chi_2+ -> ~b_2*    t )
     1.96925255E-02    2     1000012       -11   # BR(~chi_2+ -> ~nu_eL   e+  )
     1.96925255E-02    2     1000014       -13   # BR(~chi_2+ -> ~nu_muL  mu+ )
     2.66937138E-02    2     1000016       -15   # BR(~chi_2+ -> ~nu_tau1 tau+)
     5.23600711E-02    2    -1000011        12   # BR(~chi_2+ -> ~e_L+    nu_e)
     0.00000000E+00    2    -2000011        12   # BR(~chi_2+ -> ~e_R+    nu_e)
     5.23600711E-02    2    -1000013        14   # BR(~chi_2+ -> ~mu_L+   nu_mu)
     0.00000000E+00    2    -2000013        14   # BR(~chi_2+ -> ~mu_R+   nu_mu)
     2.27029940E-04    2    -1000015        16   # BR(~chi_2+ -> ~tau_1+  nu_tau)
     5.67355363E-02    2    -2000015        16   # BR(~chi_2+ -> ~tau_2+  nu_tau)
     2.42517968E-01    2     1000024        23   # BR(~chi_2+ -> ~chi_1+  Z )
     6.41934286E-02    2     1000022        24   # BR(~chi_2+ -> ~chi_10  W+)
     2.95919427E-01    2     1000023        24   # BR(~chi_2+ -> ~chi_20  W+)
     0.00000000E+00    2     1000025        24   # BR(~chi_2+ -> ~chi_30  W+)
     0.00000000E+00    2     1000035        24   # BR(~chi_2+ -> ~chi_40  W+)
     1.69607703E-01    2     1000024        25   # BR(~chi_2+ -> ~chi_1+  h )
     0.00000000E+00    2     1000024        35   # BR(~chi_2+ -> ~chi_1+  H )
     0.00000000E+00    2     1000024        36   # BR(~chi_2+ -> ~chi_1+  A )
     0.00000000E+00    2     1000022        37   # BR(~chi_2+ -> ~chi_10  H+)
     0.00000000E+00    2     1000023        37   # BR(~chi_2+ -> ~chi_20  H+)
     0.00000000E+00    2     1000025        37   # BR(~chi_2+ -> ~chi_30  H+)
     0.00000000E+00    2     1000035        37   # BR(~chi_2+ -> ~chi_40  H+)
#
#         PDG            Width
DECAY   1000022     0.00000000E+00   # neutralino1 decays
#
#         PDG            Width
DECAY   1000023     1.64735255E-02   # neutralino2 decays
#          BR         NDA      ID1       ID2
     0.00000000E+00    2     1000022        23   # BR(~chi_20 -> ~chi_10   Z )
     0.00000000E+00    2     1000024       -24   # BR(~chi_20 -> ~chi_1+   W-)
     0.00000000E+00    2    -1000024        24   # BR(~chi_20 -> ~chi_1-   W+)
     0.00000000E+00    2     1000037       -24   # BR(~chi_20 -> ~chi_2+   W-)
     0.00000000E+00    2    -1000037        24   # BR(~chi_20 -> ~chi_2-   W+)
     0.00000000E+00    2     1000022        25   # BR(~chi_20 -> ~chi_10   h )
     0.00000000E+00    2     1000022        35   # BR(~chi_20 -> ~chi_10   H )
     0.00000000E+00    2     1000022        36   # BR(~chi_20 -> ~chi_10   A )
     0.00000000E+00    2     1000024       -37   # BR(~chi_20 -> ~chi_1+   H-)
     0.00000000E+00    2    -1000024        37   # BR(~chi_20 -> ~chi_1-   H+)
     0.00000000E+00    2     1000037       -37   # BR(~chi_20 -> ~chi_2+   H-)
     0.00000000E+00    2    -1000037        37   # BR(~chi_20 -> ~chi_2-   H+)
     0.00000000E+00    2     1000002        -2   # BR(~chi_20 -> ~u_L      ub)
     0.00000000E+00    2    -1000002         2   # BR(~chi_20 -> ~u_L*     u )
     0.00000000E+00    2     2000002        -2   # BR(~chi_20 -> ~u_R      ub)
     0.00000000E+00    2    -2000002         2   # BR(~chi_20 -> ~u_R*     u )
     0.00000000E+00    2     1000001        -1   # BR(~chi_20 -> ~d_L      db)
     0.00000000E+00    2    -1000001         1   # BR(~chi_20 -> ~d_L*     d )
     0.00000000E+00    2     2000001        -1   # BR(~chi_20 -> ~d_R      db)
     0.00000000E+00    2    -2000001         1   # BR(~chi_20 -> ~d_R*     d )
     0.00000000E+00    2     1000004        -4   # BR(~chi_20 -> ~c_L      cb)
     0.00000000E+00    2    -1000004         4   # BR(~chi_20 -> ~c_L*     c )
     0.00000000E+00    2     2000004        -4   # BR(~chi_20 -> ~c_R      cb)
     0.00000000E+00    2    -2000004         4   # BR(~chi_20 -> ~c_R*     c )
     0.00000000E+00    2     1000003        -3   # BR(~chi_20 -> ~s_L      sb)
     0.00000000E+00    2    -1000003         3   # BR(~chi_20 -> ~s_L*     s )
     0.00000000E+00    2     2000003        -3   # BR(~chi_20 -> ~s_R      sb)
     0.00000000E+00    2    -2000003         3   # BR(~chi_20 -> ~s_R*     s )
     0.00000000E+00    2     1000006        -6   # BR(~chi_20 -> ~t_1      tb)
     0.00000000E+00    2    -1000006         6   # BR(~chi_20 -> ~t_1*     t )
     0.00000000E+00    2     2000006        -6   # BR(~chi_20 -> ~t_2      tb)
     0.00000000E+00    2    -2000006         6   # BR(~chi_20 -> ~t_2*     t )
     0.00000000E+00    2     1000005        -5   # BR(~chi_20 -> ~b_1      bb)
     0.00000000E+00    2    -1000005         5   # BR(~chi_20 -> ~b_1*     b )
     0.00000000E+00    2     2000005        -5   # BR(~chi_20 -> ~b_2      bb)
     0.00000000E+00    2    -2000005         5   # BR(~chi_20 -> ~b_2*     b )
     0.00000000E+00    2     1000011       -11   # BR(~chi_20 -> ~e_L-     e+)
     0.00000000E+00    2    -1000011        11   # BR(~chi_20 -> ~e_L+     e-)
     3.29566665E-02    2     2000011       -11   # BR(~chi_20 -> ~e_R-     e+)
     3.29566665E-02    2    -2000011        11   # BR(~chi_20 -> ~e_R+     e-)
     0.00000000E+00    2     1000013       -13   # BR(~chi_20 -> ~mu_L-    mu+)
     0.00000000E+00    2    -1000013        13   # BR(~chi_20 -> ~mu_L+    mu-)
     3.29566665E-02    2     2000013       -13   # BR(~chi_20 -> ~mu_R-    mu+)
     3.29566665E-02    2    -2000013        13   # BR(~chi_20 -> ~mu_R+    mu-)
     4.34086667E-01    2     1000015       -15   # BR(~chi_20 -> ~tau_1-   tau+)
     4.34086667E-01    2    -1000015        15   # BR(~chi_20 -> ~tau_1+   tau-)
     0.00000000E+00    2     2000015       -15   # BR(~chi_20 -> ~tau_2-   tau+)
     0.00000000E+00    2    -2000015        15   # BR(~chi_20 -> ~tau_2+   tau-)
     0.00000000E+00    2     1000012       -12   # BR(~chi_20 -> ~nu_eL    nu_eb)
     0.00000000E+00    2    -1000012        12   # BR(~chi_20 -> ~nu_eL*   nu_e )
     0.00000000E+00    2     1000014       -14   # BR(~chi_20 -> ~nu_muL   nu_mub)
     0.00000000E+00    2    -1000014        14   # BR(~chi_20 -> ~nu_muL*  nu_mu )
     0.00000000E+00    2     1000016       -16   # BR(~chi_20 -> ~nu_tau1  nu_taub)
     0.00000000E+00    2    -1000016        16   # BR(~chi_20 -> ~nu_tau1* nu_tau )
#
#         PDG            Width
DECAY   1000025     1.92862382E+00   # neutralino3 decays
#          BR         NDA      ID1       ID2
     1.15279015E-01    2     1000022        23   # BR(~chi_30 -> ~chi_10   Z )
     2.11120573E-01    2     1000023        23   # BR(~chi_30 -> ~chi_20   Z )
     2.97321131E-01    2     1000024       -24   # BR(~chi_30 -> ~chi_1+   W-)
     2.97321131E-01    2    -1000024        24   # BR(~chi_30 -> ~chi_1-   W+)
     0.00000000E+00    2     1000037       -24   # BR(~chi_30 -> ~chi_2+   W-)
     0.00000000E+00    2    -1000037        24   # BR(~chi_30 -> ~chi_2-   W+)
     1.87138985E-02    2     1000022        25   # BR(~chi_30 -> ~chi_10   h )
     0.00000000E+00    2     1000022        35   # BR(~chi_30 -> ~chi_10   H )
     0.00000000E+00    2     1000022        36   # BR(~chi_30 -> ~chi_10   A )
     1.14519963E-02    2     1000023        25   # BR(~chi_30 -> ~chi_20   h )
     0.00000000E+00    2     1000023        35   # BR(~chi_30 -> ~chi_20   H )
     0.00000000E+00    2     1000023        36   # BR(~chi_30 -> ~chi_20   A )
     0.00000000E+00    2     1000024       -37   # BR(~chi_30 -> ~chi_1+   H-)
     0.00000000E+00    2    -1000024        37   # BR(~chi_30 -> ~chi_1-   H+)
     0.00000000E+00    2     1000037       -37   # BR(~chi_30 -> ~chi_2+   H-)
     0.00000000E+00    2    -1000037        37   # BR(~chi_30 -> ~chi_2-   H+)
     0.00000000E+00    2     1000002        -2   # BR(~chi_30 -> ~u_L      ub)
     0.00000000E+00    2    -1000002         2   # BR(~chi_30 -> ~u_L*     u )
     0.00000000E+00    2     2000002        -2   # BR(~chi_30 -> ~u_R      ub)
     0.00000000E+00    2    -2000002         2   # BR(~chi_30 -> ~u_R*     u )
     0.00000000E+00    2     1000001        -1   # BR(~chi_30 -> ~d_L      db)
     0.00000000E+00    2    -1000001         1   # BR(~chi_30 -> ~d_L*     d )
     0.00000000E+00    2     2000001        -1   # BR(~chi_30 -> ~d_R      db)
     0.00000000E+00    2    -2000001         1   # BR(~chi_30 -> ~d_R*     d )
     0.00000000E+00    2     1000004        -4   # BR(~chi_30 -> ~c_L      cb)
     0.00000000E+00    2    -1000004         4   # BR(~chi_30 -> ~c_L*     c )
     0.00000000E+00    2     2000004        -4   # BR(~chi_30 -> ~c_R      cb)
     0.00000000E+00    2    -2000004         4   # BR(~chi_30 -> ~c_R*     c )
     0.00000000E+00    2     1000003        -3   # BR(~chi_30 -> ~s_L      sb)
     0.00000000E+00    2    -1000003         3   # BR(~chi_30 -> ~s_L*     s )
     0.00000000E+00    2     2000003        -3   # BR(~chi_30 -> ~s_R      sb)
     0.00000000E+00    2    -2000003         3   # BR(~chi_30 -> ~s_R*     s )
     0.00000000E+00    2     1000006        -6   # BR(~chi_30 -> ~t_1      tb)
     0.00000000E+00    2    -1000006         6   # BR(~chi_30 -> ~t_1*     t )
     0.00000000E+00    2     2000006        -6   # BR(~chi_30 -> ~t_2      tb)
     0.00000000E+00    2    -2000006         6   # BR(~chi_30 -> ~t_2*     t )
     0.00000000E+00    2     1000005        -5   # BR(~chi_30 -> ~b_1      bb)
     0.00000000E+00    2    -1000005         5   # BR(~chi_30 -> ~b_1*     b )
     0.00000000E+00    2     2000005        -5   # BR(~chi_30 -> ~b_2      bb)
     0.00000000E+00    2    -2000005         5   # BR(~chi_30 -> ~b_2*     b )
     6.15727811E-04    2     1000011       -11   # BR(~chi_30 -> ~e_L-     e+)
     6.15727811E-04    2    -1000011        11   # BR(~chi_30 -> ~e_L+     e-)
     1.16411010E-03    2     2000011       -11   # BR(~chi_30 -> ~e_R-     e+)
     1.16411010E-03    2    -2000011        11   # BR(~chi_30 -> ~e_R+     e-)
     6.15727811E-04    2     1000013       -13   # BR(~chi_30 -> ~mu_L-    mu+)
     6.15727811E-04    2    -1000013        13   # BR(~chi_30 -> ~mu_L+    mu-)
     1.16411010E-03    2     2000013       -13   # BR(~chi_30 -> ~mu_R-    mu+)
     1.16411010E-03    2    -2000013        13   # BR(~chi_30 -> ~mu_R+    mu-)
     4.92152638E-03    2     1000015       -15   # BR(~chi_30 -> ~tau_1-   tau+)
     4.92152638E-03    2    -1000015        15   # BR(~chi_30 -> ~tau_1+   tau-)
     6.41442155E-03    2     2000015       -15   # BR(~chi_30 -> ~tau_2-   tau+)
     6.41442155E-03    2    -2000015        15   # BR(~chi_30 -> ~tau_2+   tau-)
     3.15943787E-03    2     1000012       -12   # BR(~chi_30 -> ~nu_eL    nu_eb)
     3.15943787E-03    2    -1000012        12   # BR(~chi_30 -> ~nu_eL*   nu_e )
     3.15943787E-03    2     1000014       -14   # BR(~chi_30 -> ~nu_muL   nu_mub)
     3.15943787E-03    2    -1000014        14   # BR(~chi_30 -> ~nu_muL*  nu_mu )
     3.18162814E-03    2     1000016       -16   # BR(~chi_30 -> ~nu_tau1  nu_taub)
     3.18162814E-03    2    -1000016        16   # BR(~chi_30 -> ~nu_tau1* nu_tau )
#
#         PDG            Width
DECAY   1000035     2.63644492E+00   # neutralino4 decays
#          BR         NDA      ID1       ID2
     2.08250509E-02    2     1000022        23   # BR(~chi_40 -> ~chi_10   Z )
     1.85725667E-02    2     1000023        23   # BR(~chi_40 -> ~chi_20   Z )
     0.00000000E+00    2     1000025        23   # BR(~chi_40 -> ~chi_30   Z )
     2.61809331E-01    2     1000024       -24   # BR(~chi_40 -> ~chi_1+   W-)
     2.61809331E-01    2    -1000024        24   # BR(~chi_40 -> ~chi_1-   W+)
     0.00000000E+00    2     1000037       -24   # BR(~chi_40 -> ~chi_2+   W-)
     0.00000000E+00    2    -1000037        24   # BR(~chi_40 -> ~chi_2-   W+)
     6.29362490E-02    2     1000022        25   # BR(~chi_40 -> ~chi_10   h )
     0.00000000E+00    2     1000022        35   # BR(~chi_40 -> ~chi_10   H )
     0.00000000E+00    2     1000022        36   # BR(~chi_40 -> ~chi_10   A )
     1.33506859E-01    2     1000023        25   # BR(~chi_40 -> ~chi_20   h )
     0.00000000E+00    2     1000023        35   # BR(~chi_40 -> ~chi_20   H )
     0.00000000E+00    2     1000023        36   # BR(~chi_40 -> ~chi_20   A )
     0.00000000E+00    2     1000025        25   # BR(~chi_40 -> ~chi_30   h )
     0.00000000E+00    2     1000025        35   # BR(~chi_40 -> ~chi_30   H )
     0.00000000E+00    2     1000025        36   # BR(~chi_40 -> ~chi_30   A )
     0.00000000E+00    2     1000024       -37   # BR(~chi_40 -> ~chi_1+   H-)
     0.00000000E+00    2    -1000024        37   # BR(~chi_40 -> ~chi_1-   H+)
     0.00000000E+00    2     1000037       -37   # BR(~chi_40 -> ~chi_2+   H-)
     0.00000000E+00    2    -1000037        37   # BR(~chi_40 -> ~chi_2-   H+)
     0.00000000E+00    2     1000002        -2   # BR(~chi_40 -> ~u_L      ub)
     0.00000000E+00    2    -1000002         2   # BR(~chi_40 -> ~u_L*     u )
     0.00000000E+00    2     2000002        -2   # BR(~chi_40 -> ~u_R      ub)
     0.00000000E+00    2    -2000002         2   # BR(~chi_40 -> ~u_R*     u )
     0.00000000E+00    2     1000001        -1   # BR(~chi_40 -> ~d_L      db)
     0.00000000E+00    2    -1000001         1   # BR(~chi_40 -> ~d_L*     d )
     0.00000000E+00    2     2000001        -1   # BR(~chi_40 -> ~d_R      db)
     0.00000000E+00    2    -2000001         1   # BR(~chi_40 -> ~d_R*     d )
     0.00000000E+00    2     1000004        -4   # BR(~chi_40 -> ~c_L      cb)
     0.00000000E+00    2    -1000004         4   # BR(~chi_40 -> ~c_L*     c )
     0.00000000E+00    2     2000004        -4   # BR(~chi_40 -> ~c_R      cb)
     0.00000000E+00    2    -2000004         4   # BR(~chi_40 -> ~c_R*     c )
     0.00000000E+00    2     1000003        -3   # BR(~chi_40 -> ~s_L      sb)
     0.00000000E+00    2    -1000003         3   # BR(~chi_40 -> ~s_L*     s )
     0.00000000E+00    2     2000003        -3   # BR(~chi_40 -> ~s_R      sb)
     0.00000000E+00    2    -2000003         3   # BR(~chi_40 -> ~s_R*     s )
     0.00000000E+00    2     1000006        -6   # BR(~chi_40 -> ~t_1      tb)
     0.00000000E+00    2    -1000006         6   # BR(~chi_40 -> ~t_1*     t )
     0.00000000E+00    2     2000006        -6   # BR(~chi_40 -> ~t_2      tb)
     0.00000000E+00    2    -2000006         6   # BR(~chi_40 -> ~t_2*     t )
     0.00000000E+00    2     1000005        -5   # BR(~chi_40 -> ~b_1      bb)
     0.00000000E+00    2    -1000005         5   # BR(~chi_40 -> ~b_1*     b )
     0.00000000E+00    2     2000005        -5   # BR(~chi_40 -> ~b_2      bb)
     0.00000000E+00    2    -2000005         5   # BR(~chi_40 -> ~b_2*     b )
     9.78996815E-03    2     1000011       -11   # BR(~chi_40 -> ~e_L-     e+)
     9.78996815E-03    2    -1000011        11   # BR(~chi_40 -> ~e_L+     e-)
     3.59737855E-03    2     2000011       -11   # BR(~chi_40 -> ~e_R-     e+)
     3.59737855E-03    2    -2000011        11   # BR(~chi_40 -> ~e_R+     e-)
     9.78996815E-03    2     1000013       -13   # BR(~chi_40 -> ~mu_L-    mu+)
     9.78996815E-03    2    -1000013        13   # BR(~chi_40 -> ~mu_L+    mu-)
     3.59737855E-03    2     2000013       -13   # BR(~chi_40 -> ~mu_R-    mu+)
     3.59737855E-03    2    -2000013        13   # BR(~chi_40 -> ~mu_R+    mu-)
     2.50391561E-03    2     1000015       -15   # BR(~chi_40 -> ~tau_1-   tau+)
     2.50391561E-03    2    -1000015        15   # BR(~chi_40 -> ~tau_1+   tau-)
     1.59074552E-02    2     2000015       -15   # BR(~chi_40 -> ~tau_2-   tau+)
     1.59074552E-02    2    -2000015        15   # BR(~chi_40 -> ~tau_2+   tau-)
     2.49773159E-02    2     1000012       -12   # BR(~chi_40 -> ~nu_eL    nu_eb)
     2.49773159E-02    2    -1000012        12   # BR(~chi_40 -> ~nu_eL*   nu_e )
     2.49773159E-02    2     1000014       -14   # BR(~chi_40 -> ~nu_muL   nu_mub)
     2.49773159E-02    2    -1000014        14   # BR(~chi_40 -> ~nu_muL*  nu_mu )
     2.51296109E-02    2     1000016       -16   # BR(~chi_40 -> ~nu_tau1  nu_taub)
     2.51296109E-02    2    -1000016        16   # BR(~chi_40 -> ~nu_tau1* nu_tau )
\end{verbatim}
\subsubsection*{4.4 Some points about the program}

Our results for some representative points of the MSSM parameter space have
been carefully cross-checked against other existing codes in mSUGRA-type models
[in the case of the AMSB and GMSB models, no very detailed comparisons have
been made]. In our comparison with the programs {\tt Isajet} and {\tt
SPHENO},  when the sparticle and Higgs spectrum, as well as the various
input and soft SUSY-breaking  parameters are forced to be the same in both
codes,  we obtain in general a rather good agreement  for the partial widths
of the main two-body decay modes.  We have also verified that the loop and some
three-body decays of the top squark  agree qualitatively with those included
in, respectively, the codes  {\tt Isajet} and {\tt SPHENO}\footnote{We thank
Werner Porod for his gracious help in performing these  detailed comparisons.}.
\s

However, since there are, as is well known, some differences [of the order of a
few per cent] between the outputs of the various RGE codes [see
Ref.~\cite{comparison} for a discussion], these discrepancies  can lead to a
completely different phenomenology. Indeed, some decay  channels can be either
absent or present in some codes when the masses of the decaying and daughter
particles are close to each other. For some multibody decays and for the QCD
corrections, no comparison has been made since they are absent from these codes.
\s

The program is under rapid development and we plan to make several upgrades in
a near future. A brief and non-exhaustive list of points that will be
implemented in the next releases of the program includes: 

\begin{itemize}
\vspace*{-2mm}

\item[$(i)$] the finalization of the inclusion of the decays of the top 
quark, i.e. the implementation of the QCD corrections;  \vspace*{-2mm}

\item[$(ii)$] the link with the routine {\tt Hmasses} for a more precise and
consistent [with the program] calculation of the Higgs boson masses; 
\vspace*{-2mm}

\item[$(iii)$] the inclusion of additional higher order decay modes, such as
the bottom squark three-body decay widths and some three-body decay modes, 
which can be important in GMSB models; 
\vspace*{-2mm}

\item[$(iv)$] the inclusion of the finite widths of the propagators of the
exchanged particles in the multibody decay modes, to make smooth transitions
between the multibody and two-body decay modes; 
\vspace*{-2mm}

\item[$(v)$] at some point, the implementation of some important electroweak
radiative corrections [in particular for decays involving top and bottom
squarks]. 
\vspace*{-2mm}
\end{itemize}

As mentioned earlier, a version where the program {\tt SDECAY} is linked  not
only to the program {\tt SuSpect} but also to the  program {\tt HDECAY} for the
calculation of the decay widths and branching ratios of the MSSM Higgs bosons
is in preparation. It would allow for a complete description of the
properties of SUSY particles and MSSM Higgs bosons, prior to their production
at colliders. \s

Finally, a web page devoted to the {\tt SDECAY} program can be found at the
http address:\s

\centerline{\tt http://people.web.psi.ch/muehlleitner/SDECAY}\s

\nn It contains all the information that is needed on the program: 

\begin{itemize}
\vspace*{-2mm}

\item[--] downloading directly the various files of the program;
\vspace*{-2mm}

\item[--] short explanations of the code and how to run it;
\vspace*{-2mm}

\item[--] obtaining the complete ``users manual"  in post-script or PDF
form;
\vspace*{-2mm}

\item[--] a regularly updated list of important changes/corrected bugs in the
code; 
\vspace*{-2mm}

\item[--] a mailing list to which one can subscribe to be automatically advised
about  future {\tt SDECAY}  updates  or eventual corrections.  \vspace*{-2mm}

\end{itemize}

\subsection*{5. Conclusions} 

We have presented the Fortran code {\tt SDECAY}, which evaluates the decay
widths and branching ratios of the supersymmetric particles in the MSSM. It
includes not only all the possible tree-level decays into two-body final
states,  but also the various three-body modes for charginos, neutralinos,
gluinos and top squarks, the four-body decays of the lightest
top squark and the loop-induced decays of the lightest top squark, the
next-to-lightest neutralino and the gluino. In addition, the QCD corrections to
the tree-level two-body decays, which involve strongly interacting particles and
the dominant electroweak corrections due to the running of the gauge and
fermion Yukawa couplings are incorporated. Furthermore, we have included the
decays of the  NLSP in GMSB models and the standard and SUSY decay modes of the
heavy top quark. \s

The program uses the {\tt SuSpect} code for the calculation of the spectrum and
the various soft SUSY-breaking parameters, but it can be easily linked to any
other RGE code. It can be also linked to the program {\tt HDECAY} for the
decays of the MSSM Higgs bosons, to provide a complete picture of the new
particles predicted in the MSSM, except for their production properties. The
latter are dealt with by the Monte Carlo event generators with which {\tt
SDECAY} can be easily linked since, in particular, it generates an output \`a
la SUSY Les Houches Accord. The program is user-friendly, flexible for the
choice of options and approximations, and quite fast. It therefore allows for
a rather accurate, reliable and efficient study of the phenomenology of the MSSM
superparticles, including the possibility of large scans of the parameter
space. \s

The program is under rapid development and will be maintained regularly to  
include upgrades, improvements and potentially, corrections. Any suggestion,
comment or complaint from the potential users will be welcome.  \bigskip

\nn {\bf Acknowledgements:} We have benefited from  several discussions,
comments and help from: Ben Allanach, Nabil Ghodbane, Jean-Loic Kneur, 
Werner Porod, Peter Skands and Michael Spira. This work is supported by the 
EU under the contract  HPRN-CT-2000-00149.


\begin{thebibliography}{99}

\bibitem{MSSM1} For reviews on the MSSM, see:  H.P.~Nilles, Phys. Rep. 110
(1984) 1; R. Barbieri,  Riv. Nuov. Cim. 11 (1988) 1; R. Arnowitt and P. Nath,
Report CTP-TAMU-52-93; M. Drees and S. Martin, CLTP Report (1995) and
hep-ph/9504324; S. Martin,  hep-ph/9709356; J. Bagger, Lectures at TASI-95,
hep-ph/9604232. 
%
\bibitem{MSSM2} H. E. Haber and G. Kane, Phys. Rep. 117 (1985) 75;
J.F. Gunion and H.E. Haber, Nucl. Phys. B272 (1986) 1 and
B278 (1986) 449; (E) hep-ph/9301205. 
%
\bibitem{MSSM3} A. Djouadi, S. Rosier-Lees [conv.] et al., summary report of 
the MSSM Working Group for the ``GDR--Supersym\'etrie", hep--ph/9901246.
%
\bibitem{Houches} For recent analyses, see for intance the Proceedings  
of the Les Houches Workshops ``Physics at TeV colliders", A. Djouadi et al., 
hep-ph/0002258 (1999) and 
D. Cavalli et al., hep-ph/0203056 (2001). 
%
\bibitem{TESLA} E. Accomando, Phys. Rep. 299 (1998) 1;  
American Linear Collider Working Group (T. Abe et al.),  hep-ex/0106057; 
TESLA Technical Design Report, Part III, DESY--01--011C, hep-ph/0106315.  
%
\bibitem{NLOprod} For a summary, see the write-up of the talks given at SUSY02: 
M. Spira, hep-ph/0211145 and W. Majerotto, hep-ph/0209137.
%
\bibitem{isajet} H. Baer, F.E. Paige, S.D. Protopopescu and X. Tata, 
``ISAJET 7.48: A Monte Carlo event generator for $pp, p\bar{p}$ and $e^+ 
e^-$ reactions", hep-ph/0001086. 
%
\bibitem{suspect} A. Djouadi, J.L. Kneur and G. Moultaka, ``{\tt SuSpect}: a 
Fortran Code for the Supersymmetric and Higgs Particle Spectrum in the MSSM",  
hep-ph/0211331.
\bibitem{softsusy}  B.C. Allanach, ``SOFTSUSY: A C++ program for
calculating supersymmetric spectra", Comput. Phys. Commun. (2002) 143,
hep-ph/0104145. 
%
\bibitem{spheno} W. Porod, ``SPHENO: a program calculating SUSY Spectra,
SUSY particle decays and SUSY particle production at $e^+e^-$ colliders", 
Comput. Phys. Commun. 153 (2003) 275,  hep-ph/03011101. 
%
\bibitem{2body} See e.g: A. Bartl, H. Fraas and W. Majerotto, Nucl. Phys. B278
(1986) 1; for reviews and a complete list of references, see for instance: 
G.F. Giudice et al., Report of the ``Searches for New Physics" Working Group  in
``Physics at LEP2", Vol.~1, p.~463-524, hep-ph/9602207; S. Abel et al.,  Report
of the ``SUGRA" Working Group for ``RUNII at the Tevatron",  hep-ph/0003154. 
%
\bibitem{QCD1} 
S. Kraml, H. Eberl, A. Bartl, W. Majerotto and W. Porod,  Phys. Lett. B386
(1996) 175; A. Djouadi, W. Hollik and C. Junger, Phys.  Rev. D55 (1997) 6975.
%
\bibitem{QCD2} 
A. Arhrib, A. Djouadi, W. Hollik and C. Junger, Phys. Rev. D57 (1998) 5860; 
A.~Bartl, H. Eberl, K. Hidaka, S. Kraml,  W. Majerotto, W. Porod and Y. Yamada,
Phys. Rev. D59 (1999) 115007 and Phys. Lett. B419 (1998) 243;
A. Bartl et al., Phys. Lett. B435 (1998) 118.
%
\bibitem{QCD3} 
W. Beenakker, R. Hopker and P.M. Zerwas, Phys. Lett. B378 (1996) 159;
W.~Beenakker, R. Hopker, T. Plehn and P.M. Zerwas, Z. Phys. C75 (1997) 349.
%
\bibitem{NLOEWd} 
J. Guasch, W. Hollik and J. Sola, Phys. Lett. B437 (1998) 88 and B510 (2001) 
211; H.S. Hou et al., Phys. Rev. D65 (2002) 075019; Q. Li, L.G. Jin and C.S. 
Li, Phys. Rev. D65 (2002) 035007 and  D66 (2002) 115008.  
%
\bibitem{3body1}  
A. Bartl, W. Majerotto and W. Porod, Z.~Phys.~C64 (1994) 499 and Phys. Lett.
B465 (1999) 187;  H. Baer, C.H. Chen, M. Drees, F. Paige and  X. Tata, Phys.
Rev. D59 (1999) 055014 and Phys. Rev. Lett. 79 (1997) 986. 
%
\bibitem{3body2} 
A. Djouadi, Y. Mambrini and M. M\"uhlleitner, Eur. Phys. J. C20 (2001) 563. 
%
\bibitem{3body3} 
W. Porod and T. Wohrmann, Phys. Rev. D55 (1997) 2907; W. Porod, Phys. Rev. D59
(1999) 095009; A. Datta, M. Guchait and K.K. Jeong,  Int. J. Mod. Phys. A14 
(1999) 2239; A. Djouadi, M. Guchait and Y. Mambrini, Phys. Rev. D64 (2001) 
095014. 
%
\bibitem{3body4} 
A. Djouadi and Y. Mambrini, Phys. Lett. B493 (2000) 120 
and Phys. Rev. D63 (2001) 115005. 
%
\bibitem{4body} 
C. Boehm, A. Djouadi and Y. Mambrini, Phys. Rev. D61 (2000) 095006. 
%
\bibitem{loop1} 
K.I. Hikasa and M. Kobayashi, Phys. Rev. D36 (1987) 724. 
%
\bibitem{loop2} 
H.E. Haber and D. Wyler, Nucl. Phys. B323 (1989) 267. 
%
\bibitem{loop3} 
S. Ambrosanio and B. Mele, Phys.Rev. D53 (1996) 2541 and D55 (1997) 1399;
H.~Baer and T. Krupovnickas, JHEP 0209 (2002) 038. 
%
\bibitem{gluinorad} 
E. Ma and G.G. Wong, Mod. Phys. Lett. A3 (1988) 1561; 
R. Barbieri et al., Nucl. Phys. B301 (1988) 15;
H. Baer, X. Tata and J. Woodside, Phys. Rev. D42 (1990) 1568.
%
\bibitem{susygen} S. Katsanevas and P. Morawitz, ``SUSYGEN: A Monte Carlo
event generator for MSSM sparticle production at $e^+e^-$ colliders", 
Comput.\,Phys.\,Com.\,112\,(1998)\,227.
%
\bibitem{QCDtop} A. Dabelstein et al., Nucl. Phys. B454 (1995) 75;
J. Guasch, R. A. Jimenez and J.~Sola, Phys. Lett. B360 (1995) 47;
A. Bartl et al., Phys. Lett. B378 (1996) 167;
A. Djouadi, W. Hollik and C. Junger, Phys. Rev. D54 (1996) 5629;
C.S. Li, R.J. Oakes and J.M. Yang, Phys. Rev. D54 (1996) 6883.
%
 %
\bibitem{SLHA} P. Skands et al., ``SUSY Les Houches Accord: Interfacing SUSY
spectrum calculators, decay packages, and event generators", 
hep-ph/0311123.
%
\bibitem{hdecay} A. Djouadi, J. Kalinowski and M. Spira, ``{\tt HDECAY}: A 
program for Higgs bosons decays in the SM and its supersymmetric extension",  
Comput. Phys. Com. 108 (1998) 56.
%
\bibitem{SSH} A. Djouadi et al., in preparation. 
%
\bibitem{mSUGRA} For a recent review of mSUGRA, see: M. Drees and S. Martin 
in Ref.~\cite{MSSM1}. 
%
\bibitem{GMSB} For a general review: G.F. Giudice and R. Rattazzi, 
Phys.~Rep.~322 (1999) 419. 
%
\bibitem{AMSB} See e.g.,
L. Randall and R. Sundrum, Nucl. Phys. B557 (1999) 79;
G. Giudice, M. Luty, H. Murayama and R. Rattazzi, JHEP 9812 (1998) 027;
J.A. Bagger, T. Moroi and E. Poppitz, JHEP 0004 (2000) 009;
K. Huitu, J. Laamanen and P. N. Pandita, Phys.~Rev. D65 (2002) 115003.
%
\bibitem{PBMZ}  D.M. Pierce, J.A. Bagger, K. Matchev and R.J. Zhang, Nucl.
Phys. B491 (1997) 3. 
%
\bibitem{SUBH} M. Carena, J.\,Espinosa, M. Quiros and C.\,Wagner, Phys. 
Lett. B335 (1995) 209.
%
\bibitem{HHH} H. Haber, R. Hempfling and A. Hoang, Z. Phys. C75 (1997) 539.
%
\bibitem{FeynHiggsFast} S. Heinemeyer, W. Hollik and G. Weiglein, Comput.\,
Phys.\,Commun. 124 (2000) 76 and hep-ph/0002213.
%
\bibitem{BDSZ} Routines written by P. Slavich and based on: G. Degrassi, P. 
Slavich and F. Zwirner, Nucl. Phys. B611 (2001) 403; A. Brignole, G. Degrassi, 
P. Slavich and F. Zwirner, Nucl. Phys. B631 (2002) 195 and Nucl. Phys. B643 
(2002) 79; A. Dedes and  P. Slavich, Nucl. Phys. B657 (2003) 333. 
%
\bibitem{QCDrunning} A. Djouadi, M. Spira and P.M. Zerwas, 
Z. Phys. C70 (1996) 427. 
%
\bibitem{PV} G. Passarino and M. Veltman, Nucl. Phys. B160 (1979) 151.
%
\bibitem{porodgluino} 
W. Porod, JHEP 0205 (2002) 030, [hep-ph/0202259].
%
\bibitem{dattagluino} A. Datta, A. Djouadi, M. Guchait and Y. Mambrini,
Phys. Rev. D65 (2002) 015007.
%
\bibitem{PDG} Particle Data Group (K. Hagiwara et al)., Phys. Rev.
D66 (2002) 010001.
%
\bibitem{benchmark} B.C. Allanach et al., ``The Snowmass points and slopes: 
Benchmarks for SUSY searches", Eur. Phys. J. C25 (2002) 113 [hep-ph/0202233].
%
\bibitem{comparison}
B.\,Allanach, S.\,Kraml and W.\,Porod, JHEP 0303 (2003) 016 and talk given 
at SUSY02,  hep-ph/0207314;  A. Djouadi, talk given at SUSY02, hep-ph/0211357.

\end{thebibliography}
\end{document}